# Tunable inertia of chiral magnetic domain walls

Jacob Torrejon[1,2], Eduardo Martinez[3] and Masamitsu Hayashi[1,4*]

[1]*National Institute for Materials Science, Tsukuba 305-0047, Japan*

[2]*Unité Mixte de Physique CNRS/Thales, 1 Avenue Augustin Fresnel, 91767 Palaiseau, France*

[3]*Departamento de Fisica Aplicada, Universidad de Salamanca, Plaza de los Caidos s/n, E-37008 Salamanca, Spain*

[4]*Department of Physics, The University of Tokyo, Bunkyo, Tokyo 113-0033, Japan*

The time it takes to accelerate an object from zero to a given velocity depends on the applied force and the environment. If the force ceases, it takes exactly the same time to completely decelerate. A magnetic domain wall (DW) is a topological object that has been observed to follow this behavior. Here we show that acceleration and deceleration times of chiral Neel walls driven by current are different in a system with low damping and moderate Dzyaloshinskii-Moriya (DM) exchange constant. The time needed to accelerate a DW with current via the spin Hall torque is much faster than the time it needs to decelerate once the current is turned off. The deceleration time is defined by the DM exchange constant whereas the acceleration time depends on the spin Hall torque, enabling tunable inertia of chiral DWs. Such unique feature of chiral DWs can be utilized to move and position DWs with lower current, key to the development of storage class memory devices.

*Email: hayashi@phys.s.u-tokyo.ac.jp

**Introduction**

It is now well established that a magnetic domain wall (DW) can be considered a topological object with effective mass[1-5] and momentum. For such an object, it requires certain time to accelerate right after a stimuli is turned on and to decelerate once the stimuli is removed. According to a model used to describe DWs, the acceleration and deceleration times of a DW are defined by the same material parameters that include the Gilbert damping constant, saturation magnetization and the dimension of the magnetic wire. The acceleration and deceleration times of a DW have been found to be the same when the DW is driven by current[6] via the spin transfer torque (STT) or by magnetic field[7,8]. Under such circumstances the distance a DW travels scales with the pulse length. Experimentally, identical acceleration and deceleration times manifest itself as a pulse length independent quasi-static velocity[6,9], a measure of speed obtained in experiments by dividing the total distance the DW traveled during and after the pulse application with the pulse length.

Recent reports have shown that chiral Neel DWs[10] emerge owing to the Dzyaloshinskii-Moriya (DM) interaction at interfaces of magnetic layer and a heavy metal layer with strong spin orbit coupling[11-22]. Such chiral Neel walls can be driven[23,24] by current via the spin Hall torque that arises when spin current is generated by the spin Hall effect in the heavy metal layer and diffuses into the magnetic layer[25-27].

Here we find that the quasi-static velocity of current (i.e. spin Hall torque) driven chiral DWs increases as the current pulse length is reduced, indicating that the distance a DW travels does not scale linearly with the pulse length. The change in the quasi-static velocity with pulse length depends on the current passed along the film plane as well as the film stack. Using collective coordinate and full micromagnetic models, we show that the deceleration time is significantly longer than the acceleration time, giving rise to a driving force dependent tunable inertia.

## Results

**Pulse length dependent quasi-static domain wall velocity.** The film stack studied is Si-sub/W($d$)/Co$_{20}$Fe$_{60}$B$_{20}$(1)/MgO(2)/Ta(1) (units in nanometers). Two film sets (A and B) with nominally the same film structure are made and evaluated. The magnetic and transport properties of the two sets are slightly different (see the Methods section and Supplementary Table 1). We study wires with width ($w$) of ~5 and ~50 µm patterned from the films. An optical microscopy image of a representative ~50 µm wide wire is shown in Fig. 1(i) inset together with the definition of the coordinate axis. Positive current corresponds to current flow along +$x$. Magneto-optical Kerr microscopy is used to measure the quasi-static velocity ($v_{END}$) of the DW. Positive velocity indicates that the DW moves to +$x$. (see Supplementary Note 1 and Supplementary Figs. 1 and 2 for the pulse transmission characterisitcs of a typical device).

Figure 1(a-f) shows the wall velocity as a function of pulse amplitude for films with different $d$. The pulse length ($t_P$) is fixed to 10 ns. The DWs move along the current flow regardless of the wall type (↓↑ and ↑↓ walls). For current pulses with amplitude larger than the depinning threshold, the velocity increases with increasing pulse amplitude and eventually saturates. Such trend is consistent with the DW velocity driven by the spin Hall torque [23],

$$v = v_D / \sqrt{1 + \left(\frac{J_D}{J - J_C}\right)^2} \tag{1}$$

where $v_D = \gamma \Delta H_{DM}$ is the saturation velocity and $J_D = \alpha J H_{DM}/H_{SH}$ is the current density at which the velocity saturates. $H_{DM} = \frac{D}{\Delta M_s}$ is the DM exchange field and $H_{SH} = -\frac{\hbar \theta_{SH}}{2eM_s t_{FM}} J$ is the damping-like effective field due to the spin Hall torque. Here $\gamma$ is the gyromagnetic ratio, $e$ is the electric charge, $\hbar$ is the reduced Planck constant. $\alpha$ is the Gilbert damping constant, $M_s$ is the saturation magnetization, $\Delta$ is the DW width and $t_{FM}$ is the thickness of the magnetic layer. $\theta_{SH}$ is the spin Hall angle of the heavy metal (W) layer and $D$ is the DM exchange constant. We have added an empirical threshold current density $J_C$ to Eq. (1) in order to account for the pinning. Note that Eq. (1) does not take into account transient effects which can influence the estimation of the wall velocity[28]. However, same results are also obtained by numerical solving the one dimensional (1D) collective coordinate model[1], which naturally accounts for pinning and transient effects (see Supplementary Fig. 3).

The red solid lines in Figs. 1(a-f) show fitting of the experimental data using Eq. (1). Except for the thinnest W layer device, we find that the saturation velocity decreases when the W layer thickness ($d$) is increased. The corresponding $t_P$ dependence of $v_{END}$ for each device is plotted in Fig. 1(g-l). For the thick W underlayer films, $v_{END}$ increases with decreasing pulse length. This is particularly evident when $t_P$ is shorter than ~10-20 ns. These results show that the distance a DW travels does not linearly scale with the pulse length, which is in striking difference with the STT driven DWs[6,8,9] or current driven narrow DWs in large magnetic damping system[29,30]. In contrast, $v_{END}$ drops for shorter pulses when the thickness of W is reduced below ~3 nm. See Supplementary Figs. 4 and 5 for supporting experimental results.

**The one dimensional model of domain walls.** In order to clarify the origin of the pulse length dependent velocity, the dynamics of chiral DWs under current pulses are studied using the one dimensional (1D) collective coordinate model[1] with the spin Hall torque and the DM interaction included. The wall dynamics is described using three time-dependent variables: the wall position $q(t)$, the wall magnetization angle $\psi(t)$ and the tilting angle of the wall normal $\chi(t)$[28,31,32]: see inset of Fig. 2(a) for the definition of the angles. Typical parameters of W/CoFeB/MgO (see Fig. 2 caption) are used and here we consider only the damping-like component of the spin Hall torque[26,27]. Using micromagnetic simulations we find that the presence of any field-like torque has little impact on the relaxation times which are discussed later.

The numerically calculated temporal evolution of the wall velocity $v(t)$, the magnetization angle $\psi(t)$ and the tilting angle $\chi(t)$ under current pulses with fixed amplitude ($J = 0.5 \times 10^8$ A/cm$^2$) and length ($t_\text{P} \sim 100$ ns) are shown in Fig. 2(a-c) for an ideal wire with no pinning. Note that $v(t)$ is the instantaneous velocity at time $t$ and is different from $v_\text{END}$. Two extreme damping values, $\alpha = 0.01$ (black solid line) and $\alpha = 0.3$ (red dashed line), are used in order to illustrate the transient effects.

There are two distinct features that are characteristics of spin Hall torque driven chiral DWs. First, the acceleration time (or the rise time) and the deceleration time (or the fall time) of the wall velocity are significantly different for the low damping system (Fig. 2(a), black solid line): the acceleration time is much faster than the deceleration time. Such effect is significantly suppressed when the Gilbert damping constant is larger[29,30] (Fig. 2(a), red dashed lines). Note that the acceleration/deceleration times of the velocity are correlated with those of the wall magnetization angle $\psi(t)$, see Figs. 2(a) and 2(b).

To provide a qualitative understanding, we analytically solve the differential equations of the 1D model using a linear approximation for a rigid wall ($\chi(t) = 0$). The analytical expression of the acceleration time ($\tau_\text{A}$) and deceleration time ($\tau_\text{D}$) reads (see Supplementary Note 2):

$$\tau_\text{A} = \frac{1+\alpha^2}{\gamma\left|\alpha H_\text{K}+\frac{\pi}{2}H_\text{SH}\right|} \tag{2}$$

$$\tau_\text{D} = \frac{1+\alpha^2}{\gamma\alpha\left|-H_\text{K}+\frac{\pi}{2}H_\text{DM}\right|} \tag{3}$$

where $H_\text{K} = \frac{4t_\text{FM}M_\text{s}\log(2)}{\Delta}$ is the magneto-static anisotropy field associated with the wall[28,33]. Equations (2) and (3) explicitly show the difference of the two quantities. The acceleration time depends on the spin Hall torque $H_\text{SH}$ (and therefore the current density) whereas the deceleration time is dependent on the DM exchange field $H_\text{DM}$. In the absence of the spin Hall torque and the DM exchange field, $\tau_\text{A} = \tau_\text{D} = \frac{1+\alpha^2}{|\alpha\gamma H_\text{K}|}$, which has been derived for the STT driven DWs[6]. Note that $\tau_\text{A(D)}$ evolves during the transient process (i.e. right after the current is turned on and off) and the relaxation times here represent the corresponding values when the angle magnetization is close to Bloch ($\tau_\text{A}$) or Neel ($\tau_\text{D}$) configurations. See Supplementary Note 2 and Supplementary Figs. 6 and 7 for discussion on the linearized 1D model.

The second characteristic feature of Fig. 2(a-c) is the non-negligible drop in the wall velocity after the current pulse is turned on. Such drop in the wall velocity only occurs for the tilted DWs ($\chi(t) \neq 0$)[32]. The velocity remains constant during the current pulse application for the rigid walls ($\chi(t) = 0$): compare the black solid and blue dashed lines in Fig. 2(a). Figure 2(a) and 2(c) show that the velocity decreases while the wall tilting develops. Theoretically, it has been predicted that the time needed to saturate the wall tilting scales with the square of wire width ($w$)[32]. Thus the pulse length required to observe sizable tilting becomes much longer for wider wires. We have studied the wall velocity in wires with $w\sim5$ μm and ~50 μm to clarify contribution from the tilting (see Supplementary Note 1). For the ~5 μm wires, we find signatures of wall tilting when longer current pulses are applied (Supplementary Fig. 4). However, for the wider wires, the tilting is not evident (Supplementary Fig. 5). Using typical parameters of the system, we estimate the time it takes to observe the tilting for $w\sim50$ μm becomes much longer than the maximum pulse length used here (~100 ns). Thus contribution from the wall tilting on $v_\text{END}$ is negligible when $w\sim50$ μm.

**Determination of the acceleration and deceleration times.** Thus two different phenomena contribute to the pulse length dependent wall velocity: the inertia effect that originates from the different acceleration/deceleration times and the wall tilting effect. We first estimate the acceleration and deceleration times using Eqs. (2) and (3) to quantify the inertia effect. The magnetic properties of the films are summarized in Fig. 3. The volume averaged saturation magnetization ($M/V$) and the effective magnetic anisotropy energy ($K_{EFF}$) are plotted against $d$ in Figs. 3(a) and 3(b). Using these results we calculate the domain wall anisotropy field ($H_K$) and the wall width ($\Delta$). We use $A = 1.5 \times 10^{-6}$ erg/cm, a typical value for Fe based alloys.

In order to estimate the acceleration time $\tau_A$ (Eq. (2)), one needs to know the strength of the spin Hall effective field $H_{SH}$. Here we use the spin Hall magnetoresistance (SMR)[34-36] to estimate the spin Hall angle, which allows calculation of $H_{SH}$. Interfacial effects, such as the spin memory loss[37,38] or any Rashba-Edelstein related effects[39,40], are neglected for simplicity. First, the resistivity $\rho_N$ of the W layer is obtained by fitting a linear function to the thickness dependence of the resistance inverse $1/R_{XX} \cdot (L/w)$, where $L$ and $w$ are, respectively, the length and width of the wire used to measure the device resistance. The solid line in Fig. 4(a) shows the fitting result for film set A, which gives $\rho_N \sim 150$ μΩcm. The resistivity of the W layer for film set A is slightly higher than those reported earlier[17,41,42].

The thickness dependence of the spin Hall magnetoresistance $\Delta R_{XX}/R_{XX}^Z$ is plotted in Fig. 4(b). $\Delta R_{XX}$ is the resistance difference of the device when the magnetization of the CoFeB layer points along the film plane perpendicular to the current flow ($R_{XX}^Y$) and along the film normal ($R_{XX}^Z$), i.e. $\Delta R_{XX} = R_{XX}^Y - R_{XX}^Z$. The W thickness dependence of SMR can be fitted using the following equation[41-43]: $\frac{\Delta R_{XX}}{R_{XX}^Z} = \theta_{SH}^2 \frac{\tanh(d/\lambda_N)}{(d/\lambda_N)(1+\xi)}\left[1 - \frac{1}{\cosh(d/\lambda_N)}\right]$. $\lambda_N$ is the spin diffusion length of the heavy metal (W) layer. $\xi = \rho_N t_{FM}/\rho_{FM} d$ describes the current shunting effect into the magnetic layer ($\rho_{FM}$ is the resistivity of the magnetic layer: we use $\rho_{FM}$ ~160 μΩcm from our previous study[17]). From the fitting, we obtain $|\theta_{SH}|$~0.24 and $\lambda_N$ ~1.1 nm, similar to what has been reported previously[41,42].

The spin Hall effective field ($H_{SH}$) is calculated using the above parameters. If we assume a transparent interface, $H_{SH}$ can be estimated from the following equation[25,44]: $H_{SH} = \theta_{SH} J_N \frac{\hbar}{2eM_S t_F} \left[1 - \frac{1}{\cosh(d/\lambda_N)}\right]$. (If spin memory loss is relevant for the W/CoFeB interface, $H_{SH}$ (and consequently $\tau_A$) will be underestimated.) To calculate the current density $J_N$ that flows into the W layer, we assume two parallel conducting channels (W and CoFeB layers). Calculated $H_{SH}$ is plotted in Fig. 4(c) for ~5 μm and ~50 μm wide wires when the pulse amplitude is set to 16 V. The difference in $H_{SH}$ for wires with different widths arises due to the difference in $J_N$. For both cases, however, $H_{SH}$ decreases when $d$ is larger than ~3 nm. This is primarily due to the increase in $M_S$ for larger $d$.

To evaluate the deceleration time $\tau_D$ (Eq. (3)), we must obtain the DM offset field $H_{DM}$. To do so, first the saturation velocity $v_D$ is estimated by the fitting results shown in Figs. 1(a-f). Although the velocity is estimated using 10 ns long pulses and Eq. (1) does not consider any transient effect, we assume that it gives a good estimate of $v_D$ to the first order (see Supplementary Fig.3 for the justification). $v_D$ is plotted against $d$ in Fig. 4(d) for both ~5 μm and ~50 μm wide wires. Next the DM offset field $H_{DM}$ and the DM exchange constant $D$ are calculated using the relations described after Eq. (1) and plotted against $d$ in Figs. 4(e) and 4(f), respectively. We find $D$ of ~0.3 erg/cm$^2$ that is nearly thickness independent and $H_{DM}$ decreasing with increasing $d$ due to the change in $v_D$ and $\Delta$ with $d$ (see Refs. [17,22] for $D$ of similar heterostructures).

We now have all parameters needed to calculate $\tau_A$ and $\tau_D$. The calculated values are plotted against $d$ in Fig. 4(g). In accordance with the results from the 1D model, $\tau_D$ is much larger than $\tau_A$, giving rise to the inertia effect. Note that a significantly large spin memory loss parameter[37] will be required in order to offset the difference of $\tau_A$ with $\tau_D$. The difference of the two relaxation times, $\tau_D-\tau_A$, provides a good guide for the degree of inertia and is plotted against $d$ in Fig. 4(h). $\tau_D-\tau_A$ increases with increasing thickness, reflecting the change in $H_{DM}$ with $d$.

These results can now be compared to the pulse length dependence of the wall velocity shown in Figs. 1(g-l). For the thinner W films, we find that $v_{END}$ for shorter pulses do not increase from its long pulse limit, indicating that the inertia is not observable. This is in agreement with the $d$-dependence of $\tau_D-\tau_A$ shown in Fig. 4(h) except for the device with the thinnest W layer. We note that for even thinner W samples (results not shown in Fig. 4), the domains consist of small grain-like structures and they no longer form a uniform pattern across the device. For such films, domain walls cannot be driven by current.

**Comparison to micromagnetic simulations.** Micromagnetic simulations with realistic pinning are performed to verify the inertia effect and evaluate contribution from the wall tilting (see Supplementary Note 3 for the details). The red squares in Figs. 5(a-c) show $v_{END}$ vs. $t_P$ obtained experimentally for three pulse amplitudes applied to a ~5 μm wide wire and $d$ ~3 nm. In contrast to $v_{END}$ found in wires with $w$~50 μm (Fig. 1), $v_{END}$ shows apparent reduction at longer pulses for the narrower wires ($w$~5 μm). The black circles show $v_{END}$ computed using micromagnetic simulations. The simulations are in good agreement with the experimental results. In particular, the simulations can also account for the reduction of $v_{END}$ at longer pulses ($t_P \gtrsim 20$ ns): the wall tilting effect becomes evident since the time scale for developing the tilting is close to the pulse length used when $w$~5 μm. Note that the 1D model fails to reproduce experimental results at longer pulses as it tends to underestimate the degree of wall tilting. Thus for longer pulses, where the tilting becomes more significant, the velocity reduction is larger for full micromagnetic simulations (see the Supplementary Figs. 8-10).

Figures 5(d-f) show the computed average distance ($d_{OFF}$) the DW travels after the current pulse is turned off as a function of $t_P$. $d_{OFF}$ is larger when the pulse length becomes shorter, verifying the inertia effect. Experimentally, we can estimate $d_{OFF}$ using the following relation: $d_{OFF} \sim v(t_P) \cdot \tau_D$. $v(t_P)$ is the instantaneous velocity right before the current pulse is turned off; here we assume it is close to the long pulse limit of $v_{END}$. From the results shown in Figs. 1(g-l) and Fig. 4(g), $d_{OFF}$ is in the range of ~80 nm to ~160 nm. This is in good agreement with the results from micromagnetic simulations (Fig. 5(d-f)).

**Discussion**

Although the results presented in Figs. 4(g,h) indicate that the inertia effect describes the pulse length dependence of $v_{END}$ well, other effects can influence the results. In particular, pinning is not included in deriving the relaxation times $\tau_A$ and $\tau_D$ (Eqs. (2) and (3)) and its influence can be significant in certain occasions. For example, one can imagine that the distance the wall travels after the current pulse is turned off ($d_{OFF}$) will be reduced if the pinning strength becomes significantly larger. Such effect has been observed in micromagnetic simulations and experiments in certain systems[30].

To study if there is any correlation between the degree of inertia and pinning, the average propagation field $H_P$ vs. $d$ is shown in Fig. 4(i) for the ~5 μm and ~50 μm wide wires. We find that $H_P$ takes a minimum when the domain wall width Δ is the smallest. Note that it is not always the case that $H_P$ scales with Δ. If pinning plays a dominant role in defining the inertia, we expect to see an inverse relationship between $\tau_D - \tau_A$ and $H_P$. Interestingly, this is not the case here, suggesting that the pinning is not strong enough to influence the inertia significantly.

Finally, Eqs. (2) and (3) and the numerically computed results of the 1D model (see Supplementary Fig. 7) indicate that the DW inertia significantly increases when $\frac{\pi}{2}H_{DM}$ approaches $H_K$. This is similar to what was found previously in a different system in which the inertia (i.e. the wall mass) increases when $H_K$ approaches zero as the DW makes a transition from a Neel wall to a Bloch wall[1]. Our results demonstrate that one can tune the inertia by material design, wire dimensions and, in some cases, the size of the driving force (e.g. current pulses). Large inertia can possibly lead to lower drive current for moving domain walls from pinning sites if one makes use of resonant excitation of domain walls[3]. It is possible to tune the DM interaction in such a way that inertia becomes extremely large or small. These results highlight the unique feature of current driven chiral domain walls.

**Methods**

**Sample preparation**. Films consisting of Sub./W($d$)/Co$_{20}$Fe$_{60}$B$_{20}$(1)/MgO(2)/Ta(1) (units in nanometers) are grown by magnetron sputtering on Si substrates coated with 100 nm thick SiO$_2$. Films are annealed at 300 °C ex-situ after deposition. Two film sets with nominally the same film structure are made using different sputtering systems. Magnetic and transport properties are slightly different between the two sets. A comparison of the film properties are listed in Supplementary Table 1. Wires, ~5 μm or ~50 μm wide and ~30 μm to ~40 μm long, are patterned using optical lithography and Ar ion etching. A subsequent lift-off process is used to form electrical contacts made of 5 nm Ta|100 nm Au.

**Chararcterization of the magnetic properties.** Volume averaged saturation magnetization $M/V$ and magnetic anisotropy energy $K_{EFF}$ of the films are measured using vibrating sample magnetometer (VSM). *M/V* is obtained by dividing the measured magnetic moment (*M*) by the nominal volume of the magnetic layer (*V*). The nominal volume is equal to the product of the film area (*Area*) and the thickness ($t_{FM}$) of the magnetic layer, $V = Area \cdot t_{FM}$. If a magnetic dead layer exists within the magnetic layer, *M/V* underestimates the saturation magnetization. For simplicity, here we use *M/V* for $M_S$ to estimate other quantities. The magnetic easy axis of the films points along the film normal owing to the perpendicular magnetic anisotropy originating from the CoFeB|MgO interface.

**Kerr microscopy imaging.** Motion of domain walls is studied using magneto-optical Kerr microscopy. A voltage controlled pulse generator (Picosecond Pulse Lab, model 10300B) is connected to the device. A pulse or a pulse train consisting of multiple pulses (with fixed pulse length) separated by ~10 ms is applied to the wire. Before and after the pulse(s) application, Kerr images are captured to determine the distance the domain wall traveled. The bandwidth of the cables and contact probes are DC-40 GHz. Signal transmission is limited by the pulse generator which generates a pulse with ~0.3 ns rise time and ~0.75 ns fall time.

**Domain wall velocity in wider wires.** To calculate $v_{END}$ from the Kerr images of the wider (~50 μm wide) wires, 3 to 4 rectangular sections, each ~4 μm wide, are defined. The velocity of the wall segment (↓↑ walls and ↑↓ walls) within each section is analyzed. The average $v_{END}$ of all sections is shown. Error bars denote standard deviation of $v_{END}$ for all sections (See Supplementary Note 1 and Supplementary Fig. 5). For the narrower wires (~5 μm wide) we use one section to calculate $v_{END}$.

**Data availability**

The authors declare that all data supporting the findings of this study are available within the paper and its supplementary information files.

**Acknowledgements**

This work was partly supported by MEXT R & D Next-Generation Information Technology , MEXT Grant-in-Aid for Young Scientists A (23506017) and the Grant-in-Aid for Specially Promoted Research (15H05702). The work by E. M. was supported by project WALL, FP7-PEOPLE-2013-ITN 608031 from European Commission, project MAT2014-52477-C5-4-P from Spanish government, and project SA282U14 from Junta de Castilla y Leon.

**Author contribution**

J.T. and M.H. performed the experiments. E.M. performed the micromagnetic simulations. All authors carried out the collective coordinate analysis, contributed to the analyses of the experimental data and writing the manuscript.

**Competing financial interests**

The authors declare that they have no competing financial interests.

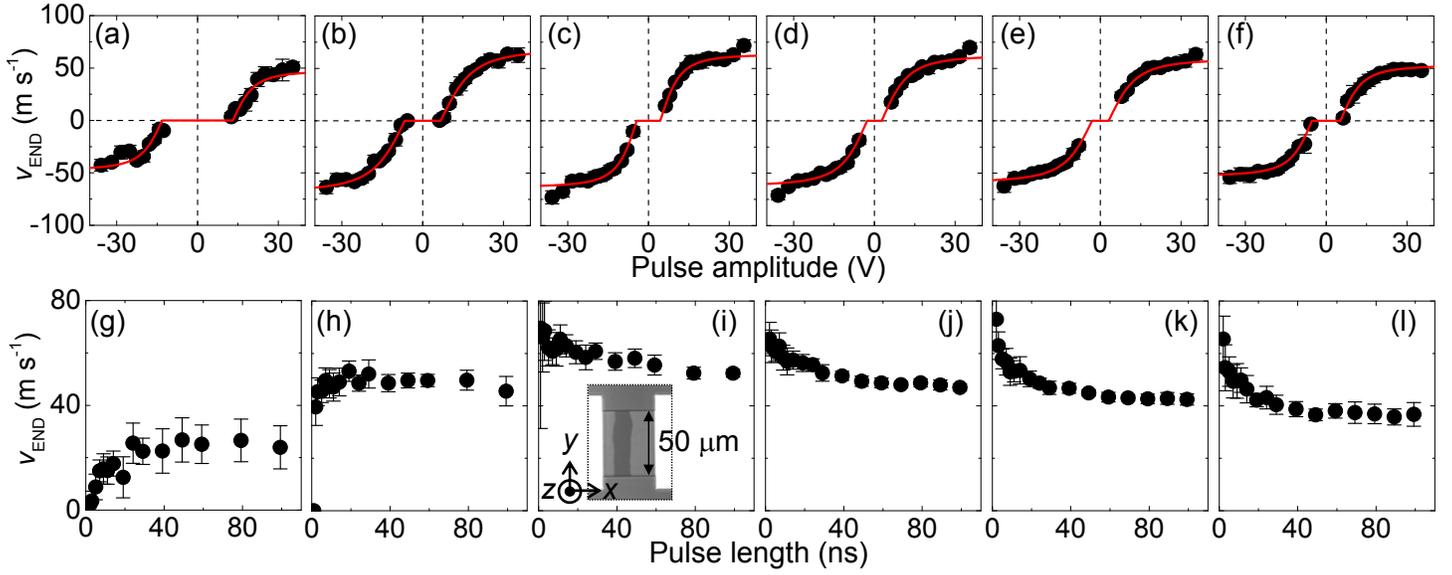

**Fig. 1. Pulse amplitude and pulse length dependent domain wall (DW) velocity.** (a-f) Quasi-static DW velocity $v_{END}$ plotted against pulse amplitude for a fixed pulse length ($t_P = 10$ ns). The red solid line represents fitting with the 1D model (Eq. (1)). (g-l) Pulse length dependence of $v_{END}$ for fixed pulse amplitude ($\pm 16$ V). Symbols represent the average $|v_{END}|$ for both positive (16 V) and negative (-16 V) pulse amplitudes. The W layer thickness $d$ varies for (a) to (f) and (g) to (l) as 2.3, 2.6, 3.0, 3.3, 3.6, 4.0 nm. Inset of (i): representative optical (Kerr) microscopy image of the device and the definition of the coordinate axis. All results are from film set A, wire width is ~50 μm. The error bars represent standard deviation of the velocity estimated in three sections of the wire (see Methods for the definition of the sections).

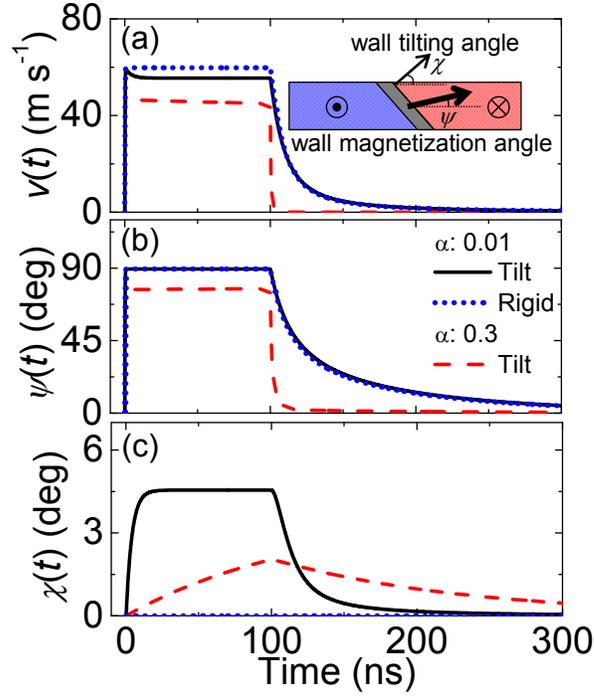

**Fig. 2. One dimensional model calculations of domain wall (DW) velocity for wires without pinning.** (a-c) Instantaneous DW velocity $v(t)$ (a), wall magnetization angle $\psi(t)$ (b) and wall tilting angle $\chi(t)$ (c) for a fixed current density ($J = 0.5 \times 10^8$ A cm$^{-2}$) flowing through the heavy metal layer. The current pulse length is ($t_P$) is 100 ns. Definition of the angles $\psi(t)$ and $\chi(t)$ are illustrated in the inset of (a). Numerical results for the rigid wall, i.e. $\chi(t) = 0$, with low damping ($\alpha = 0.01$) are shown by the blue dotted line whereas results for the tilted walls ($\chi(t) \neq 0$) are shown by the black solid ($\alpha = 0.01$) and red dashed ($\alpha = 0.3$) lines. Parameters used: saturation magnetization $M_S = 1100$ emu cm$^{-3}$, magnetic anisotropy energy $K_{EFF} = 3.2 \times 10^6$ erg cm$^{-3}$, wall width parameter $\Delta = \sqrt{A/K_{EFF}} \sim 6.8$ nm (exchange constant $A = 1.5 \times 10^{-6}$ erg cm$^{-1}$), spin Hall angle $\theta_{SH} = -0.21$, DM exchange constant $D = 0.24$ erg cm$^{-2}$, Gilbert damping constant $\alpha = 0.05$ and wire width $w$=5 μm.

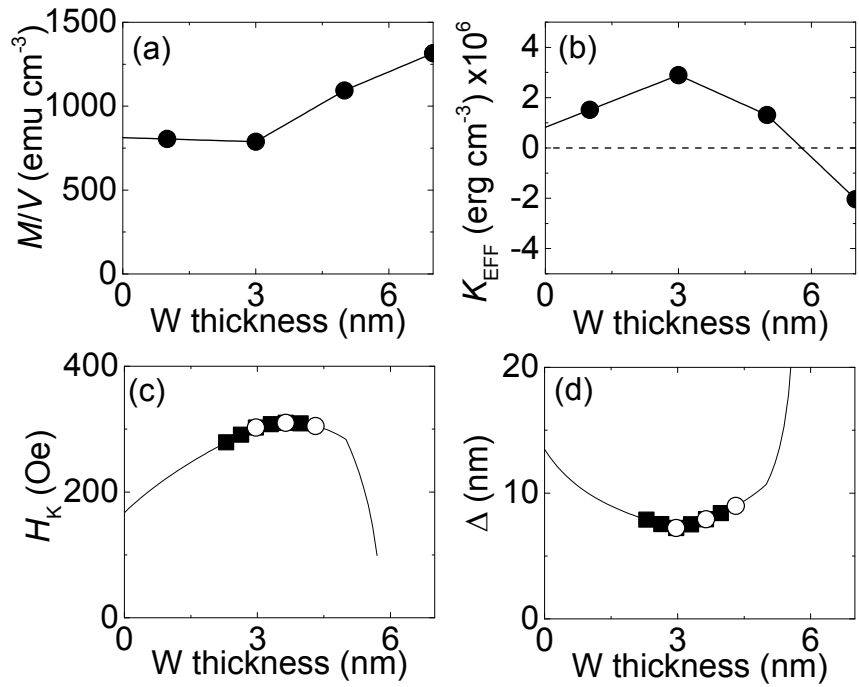

**Fig. 3. Magnetic properties of the films.** (a,b) W thickness dependence of the volume averaged saturation magnetization $M/V$ (a) and the magnetic anisotropy energy $K_{EFF}$ (b). The solid line shows linear interpolation of the data. (c,d) DW anisotropy field $H_K$ (c) and the wall width parameter $\Delta$ (d) calculated from the interpolated data shown in (a) and (b). The symbols represent values of $H_K$ and $\Delta$ that are used in the calculations presented in Fig. 4 for 5 μm wide wires (open circles) and for 50 μm wide wires (solid square). All results are from film set A.

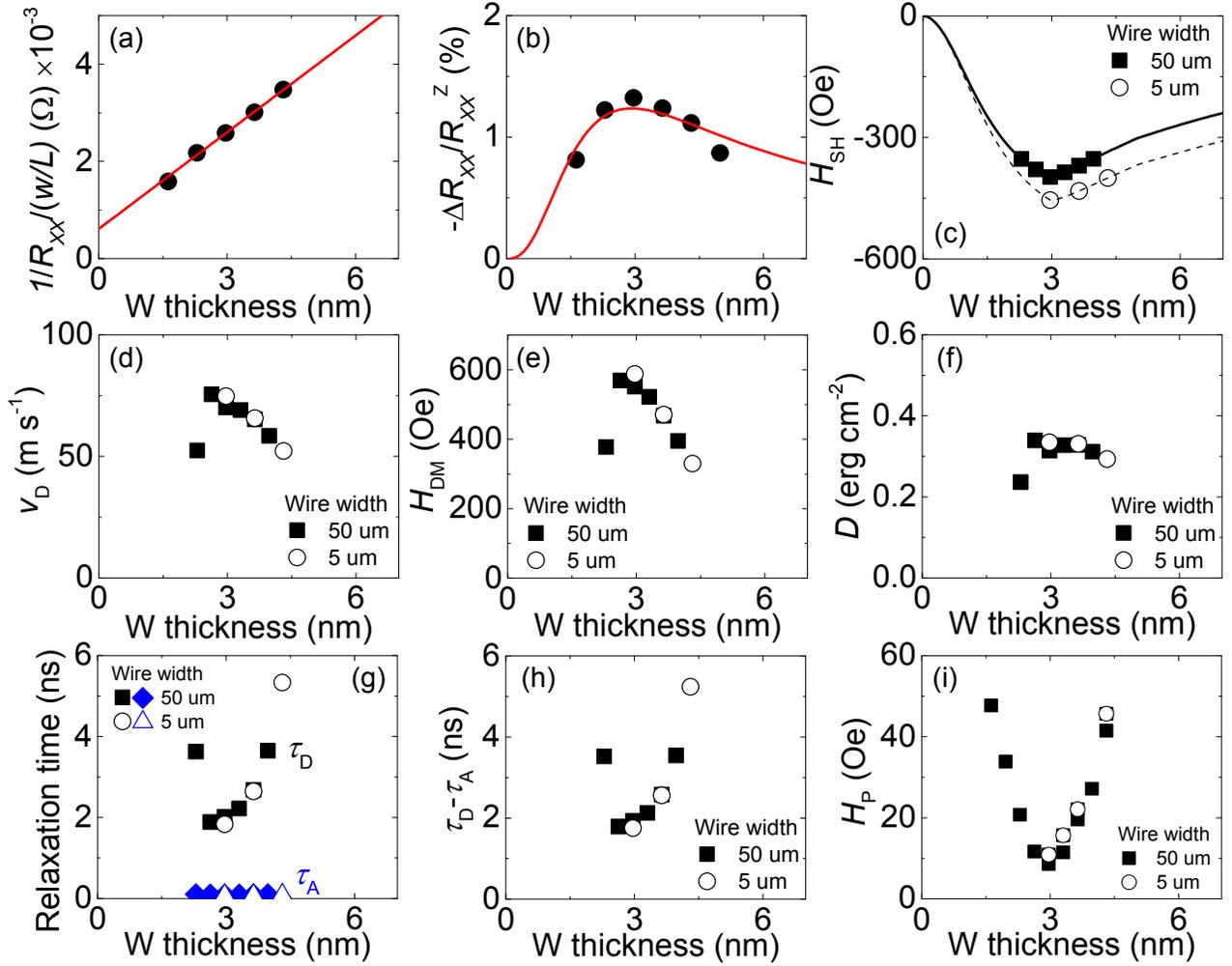

**Fig. 4. Estimated acceleration and deceleration times of domain wall (DW) motion.** (a) Normalized inverse resistance ($1/R_{XX}$) as a function of the W layer thickness. $w$ and $L$ corresponds to the width and length of the wire. Data are fitted with a linear function (solid line) to obtain the resistivity of W. (b) Spin Hall magnetoresistance $\Delta R_{XX}/R_{XX}^Z$ vs. W layer thickness. The solid line shows the fitting result. (c) Spin Hall effective field $H_{SH}$ calculated using the solid line shown in (b) when a pulse with amplitude of 16 V is applied to the wire. The solid and dashed lines display $H_{SH}$ for ~5 μm and ~50 μm wide wires, respectively. The symbols represent values of $H_{SH}$ used to calculate the acceleration time shown in (g). (d) The saturation DW velocity ($v_D$) estimated from fitting results of $v_{END}$ vs. pulse amplitude with Equation (1).. (e,f) Calculated DM offset field $H_{DM}$ (e) and the DM exchange constant $D$ (f). (g) W thickness dependence of the acceleration time ($\tau_A$) and the deceleration time ($\tau_D$) estimated using Eq. (2) and (3), respectively. Black squares (circles): $\tau_D$ for 50 μm (5 μm) wide wires, blue diamonds (triangles): $\tau_A$ for 50 μm (5 μm) wide wires. (h) Difference of $\tau_D$ and $\tau_A$ plotted against the W layer thickness. (i) Average DW propagation field $H_P$ for ~5 μm and ~50 μm wide wires plotted against the W layer thickness. All results are from film set A.

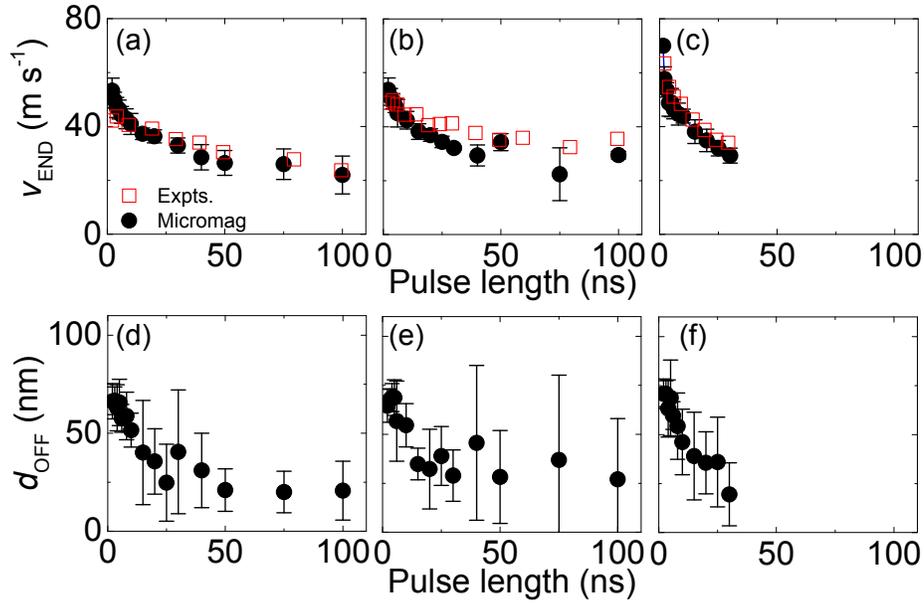

**Fig. 5. Comparison of experiments and micromagnetic simulations** (a-c) Quasi-static domain wall (DW) velocity $v_{END}$ (red squares) measured as a function of pulse length for three different pulse amplitudes: (a) 16 V ($J \sim 0.5 \times 10^8$ A cm$^{-2}$) (b) 20 V ($J \sim 0.6 \times 10^8$ A cm$^{-2}$) and (c) 25 V ($J \sim 0.8 \times 10^8$ A cm$^{-2}$). All results are from film set B, wire width is ~5 μm and the W layer thickness $d$ is ~3 nm. The black circles show calculated $v_{END}$ using micromagnetic simulations with two dimensional pinning. The average velocity is obtained from 5 different randomly generated grain patterns. (d-f) Average distance DWs travel after the current pulse is turned off ($d_{OFF}$) calculated using micromagnetic simulations. The error bars indicate distribution due to different grain patterns used in the simulations.

**Supplementary Note 1**

**Experimental setup**

*Shape of the current pulse*

Since short voltage pulses (a few nanoseconds long) are used, any glitches in the transmission line can distort the pulse shape. We thus use time domain reflection measurements to study the temporal evolution of the current that flows into the wire. Supplementary Fig. 1(a) shows schematic illustration of the measurement setup. A constant amplitude voltage pulse is applied from the pulse generator and we measure the reflected voltage pulse, via a power divider, using a real time oscilloscope. The normalized current pulses measured for pulse lengths of ~100 ns and ~2.1 ns are shown in Supplementary Fig. 1(b) and 1(c), respectively. As evident, there is no obvious glitch in the current pulse shape for both pulses. Since the input impedance of the pulse generator is not perfectly 50 Ohm, we take this into account to calculate the current applied to the device. The fluctuations in the signals found at times of ~30 ns and ~100 ns in Supplementary Fig. 1(b) are due to this correction. The difference in the rise and fall times of the pulse (Fig. S1(c)) is largely to do with the pulse generator: the rise and fall times of the pulse generator is <0.3 ns and 0.75 ns, respectively. The pulse length is measured using a real time oscilloscope.

*Effect of Joule heating*

The device temperature evolution with the current pulse due to Joule heating in a similar structure was reported previously in Ref. 1. The temperature variation was analyzed from the anomalous hall resistance. Based on these results, the temperature rise due to Joule heating is expected to be at most ~100 K for the maximum pulse amplitude and length applied to the device in this work. The increase of temperature is smaller for shorter current pulses. The effect of temperature on the pulse length dependent velocity is evaluated using micromagnetic simulations (see Supplementary Note 3, below).

Since the pulse generator outputs a constant-amplitude voltage pulse, the time evolution of the current that flows into the device may vary if Joule heating takes place. Fortunately, for the heterostructure studied here (W/CoFeB/MgO), the temperature variation of the device resistance is nearly constant due to the amorphous-like structure[2] of the conducting path (CoFeB and W). Supplementary Fig. 2 shows the measurement temperature dependence of the longitudinal resistance ($R_{XX}$) normalized by the ratio of wire width ($w$) and wire length ($L$). Unlike typical metals, the temperature variation of $R_{XX}/(w/L)$ is flat: the slope is ~−0.013 Ω/K. Thus the shape of the current pulse will not be distorted if Joule heating was to occur.

Note that the wall velocity is almost constant with the current density once it saturates; see Fig. 1(a-f) and Supplementary Fig. 3(b). Such saturation of velocity at high current is in accordance with the 1D model (Eq. (1)), which assumes constant magnetic properties at all currents. We thus consider Joule heating at high current has little impact on the magnetic properties of the films.

*Domain wall tilting in 5 μm and 50 μm wide wires*

As described in the main text, the degree of wall tilting depends on the wire width. In the following, we show results of current induced domain wall tilting and its influence on the velocity in ~5 μm and ~50 μm wide wires.

An optical microscopy image of the ~5 μm wide wire is shown in Supplementary Fig. 3(a). Supplementary Fig. 3(b), circles and squares, show measured $v_{END}$ as a function of pulse amplitude for a fixed pulse length ($t_P$ = 10 ns). The corresponding pulse length ($t_P$) dependence of $v_{END}$ is plotted in Supplementary Fig. 4(a-c), circles, for different pulse amplitudes. (The results are the same with those shown in Fig. 5(a-c).) In all cases, the velocity increases with decreasing pulse length. Signatures of domain nucleation are found for $t_P$>30 ns when the pulse amplitude is set to ~25 V (Supplementary Fig. 4(c)), which hinders accurate evaluation of the velocity in this regime. In contrast to the pulse length dependent $v_{END}$ found for the wider wires (~50 μm wide, Fig. 1(g-l)), $v_{END}$ of the narrower wires (~5 μm wide, Supplementary Fig. 4(a-c)) continues to decrease as the pulse length is increased beyond ~20 ns.

Supplementary Fig. 4(d-g) show sequences of Kerr images when voltage pulses (~16 V) are applied to the ~5 μm wide wire. The bright and dark contrasts correspond to magnetization pointing along +z and −z, respectively. The top image shows the initial state of the wire in which two domain walls with opposite wall types (↓↑ and ↑↓ walls) are placed. For $t_P$ =100 ns, the domain wall becomes tilted as it moves along the wire. We define the wall tilt angle $\chi$ as the angle between the wall normal and +x (see Fig. 2(a)). The tilt angle is opposite for the ↓↑ and ↑↓ domain walls and it reverses when the current direction is changed. The way the domain wall tilts is opposite to that if the Oersted field was to tilt the wall. This is in agreement with previous reports, which attribute the DMI as the source of the wall tilting[3-6]. For shorter pulses, the tilting is not obvious from the images.

As discussed in the main text, the wall tilting can influence the wall velocity, in particular, for longer pulses. According to Supplementary Eq. (9), the time it takes for the wall tilting to develop, defined as $\tau_\chi$, scales with the square of the wire width. Thus an increase in the wire width by a factor of ten will increase $\tau_\chi$ by 100. For typical material parameters found in this system, we expect negligible tilting when we increase the width from ~5 μm to ~50 μm for the maximum pulse length (~100 ns) used in the experiments.

To study the wire width dependence of the wall tilting, Supplementary Fig. 5 summarizes the wall motion in ~50 μm wide wire in comparison to that shown in Supplementary Fig. 4.

Supplementary Fig. 5(a) shows a Kerr microscopy image of a typical ~50 µm wide wire. Successive Kerr images of the magnetic state of the wide wires after application of current pulses are shown in Supplementary Fig. 5(c-e) for 4 ns, 20 ns and 90 ns long pulses (the pulse amplitude is fixed to ±16 V). We do not find observable wall tilting in these wide wires, in contrast to the 5 µm wide wires. However, domain walls tend to be more distorted when longer pulses (~90 ns) are used, Supplementary Fig. 5(e).

To estimate $v_{END}$, here we divide the wire into small sections and calculate the velocity of wall segments present within each section and take the average of all sections (as described in the Methods section of the main text). The yellow rectangles depicted in Supplementary Fig. 5(e), bottom right panel, show examples of the small sections. The pulse length dependence of $v_{END}$ of the wall segments present in sections A, B and C are displayed in Supplementary Figs. 5(f), 5(g) and 5(h), respectively. The black square and red circles indicate $v_{END}$ for ↑↓ and ↓↑ walls, respectively. The pulse length dependence of $v_{END}$ is similar for all segments of the walls despite the different pinning profile each segment will experience when moving along each section. The velocity obtained from the three sections are averaged and shown in Supplementary Fig. 5(b). As observed for the ~5 µm wide wires (Fig. 5(a-c) and Supplementary Fig. 4(a-c)), the velocity increases for shorter pulses ($t_P \lesssim 20$ ns). However, the gradual reduction in $v_{END}$ at longer pulses is not evident for the ~50 µm wide wires. These results indicate that the tilting effect is small for the ~50 µm wide wires, suggesting that the increase in the velocity for shorter pulses is likely to do with the inertia effect.

**Supplementary Note 2**

**The one dimensional (1D) collective coordinate model of a domain wall**

*Model description*

In order to describe the experimental observations, first the dynamics of chiral domain walls under current pulses is studied using the 1D collective coordinate model with the spin Hall torque and the DMI included[1,7-9]. The domain wall dynamics is described using the following three time ($t$) dependent variables[4,10]: the wall position $q(t)$, the wall magnetization angle $\psi(t)$ and the wall tilting angle $\chi(t)$ ($\psi(t)$ and $\chi(t)$ are defined with respect to +x, see inset of Fig. 2(a)). The tilting of the wall arises due to the DMI. Details of the 1D model used here can be found in Ref. 5.

$$(1+\alpha^2)\frac{\cos\chi}{\Delta}\frac{dq}{dt} = \left[-\frac{\gamma H_K}{2}\sin2(\psi-\chi) + \Gamma\frac{\pi}{2}\gamma H_{DM}\sin(\psi-\chi)\right] + \alpha\left[\gamma H_{PIN}(q) + \Gamma\frac{\pi}{2}\gamma H_{SH}\cos\psi\right] \quad (1)$$

$$(1+\alpha^2)\frac{d\psi}{dt} = -\alpha\left[-\frac{\gamma H_K}{2}\sin2(\psi-\chi) + \frac{\pi}{2}\Gamma\gamma H_{DM}\sin(\psi-\chi)\right] + \left[\gamma H_{PIN}(q) + \Gamma\frac{\pi}{2}\gamma H_{SH}\cos\psi\right] \quad (2)$$

$$\alpha\frac{\pi^2}{12\gamma}\left[\tan^2\chi + \left(\frac{w}{\pi\Delta\cos\chi}\right)^2\right]\frac{d\chi}{dt} = -\frac{H_K}{2}\sin(2(\psi-\chi)) + \Gamma\frac{\pi}{2}H_{DM}\sin(\psi-\chi) - \left[\frac{2K_{EFF}}{M_S} - \Gamma\frac{\pi}{2}H_{DM}\cos(\psi-\chi) + H_K\cos^2(\psi-\chi)\right]\tan\chi \quad (3)$$

For out of plane magnetized systems, the domain wall magnetization is pointing along the film plane: $\psi = 0, \pi$ and $\psi = \frac{\pi}{2}, -\frac{\pi}{2}$ corresponds to Neel and Bloch walls, respectively. The magneto-static anisotropy field associated with the wall is expressed as $H_K = \frac{4t_{FM}M_S\log(2)}{\Delta}$ [5,11], where $M_S$ is the saturation magnetization, $\Delta$ is the domain wall width parameter (the physical domain wall width is $\pi\Delta$) and $t_{FM}$ is the thickness of the magnetic layer. $\alpha$ and $K_{EFF}$ are the Gilbert damping parameter and the effective magnetic anisotropy energy of the magnetic layer, respectively. $w$ is the width of the wire. $\Gamma$ represents the domain wall pattern; $\Gamma = +1$ for the ↑↓ wall and $\Gamma = -1$ for the ↓↑ wall. $H_{PIN}(q)$, $H_{DM}$ and $H_{SH}$ are the pinning field, the Dzyalonshinskii-Moriya (DM) offset field and the spin Hall effective field, respectively. $H_{DM}$ and $H_{SH}$ can be explicitly written as $H_{DM} = \frac{D}{\Delta M_S}\Gamma$ and $H_{SH} = -\frac{\hbar \theta_{SH}}{2eM_S t_{FM}} J$ [12,13], where $D$ is the DM exchange constant, $\theta_{SH}$ is the spin Hall angle of the heavy metal layer and $J$ is the current density that flows into the heavy metal layer. Here, only the damping-like component of the spin Hall torque[13] is included. For simplicity, the spin transfer torques, both the adiabatic and the non-adiabatic terms[14], that occur within the magnetic layer is neglected since their contribution is much smaller than that of the damping-like spin Hall torque for the system under consideration[1]. The definitions of the constants used here are: $\gamma$ is the Gyromagnetic ratio, $\hbar$ is the reduced Planck constant and $e$ is the electron charge.

One can linearize Supplementary Eqs. (1) and (2) to obtain the characteristic equation of a domain wall (the wall tilting is set zero here)[1,8,9,15].

$$m\frac{\partial^2 q}{\partial t^2} + \frac{m}{\tau}\frac{\partial q}{\partial t} = F, \qquad (4)$$

where $m$ is the effective domain wall mass, $\tau$ is the relaxation time and $F$ is the driving force. These parameters are derived as:

$$m = \left(\frac{2M_S}{\gamma}\right)^2 \frac{(1+\alpha^2)wt_{FM}}{f(\psi_{eq})} \qquad (5)$$

$$\tau = \frac{\Delta}{\alpha}\left(\frac{2M_S}{\gamma}\right)\frac{(1+\alpha^2)}{f(\psi_{eq})} \qquad (6)$$

$$F = \frac{1}{f(\psi_{eq})} \left\{ \begin{array}{c} \left[ -\frac{\partial \sigma}{\partial q} + \left(\frac{2M_S}{\gamma}\right)^2 \frac{\beta u}{\Delta} \right] \cdot \frac{\partial^2 \sigma}{\partial \psi^2}\bigg|_{eq} + \Gamma \frac{\pi}{2} \gamma \sin\psi_{eq} \left[ -\left(\frac{2M_S}{\gamma}\right)^2 H_{SH} u + \left(\frac{2M_S}{\gamma}\right) H_{SH} \frac{\partial \sigma}{\partial \psi}\bigg|_{eq} \right] \\ + \Gamma \frac{\pi}{2} \gamma \cos\psi_{eq} \left[ -\left(\frac{2M_S}{\gamma}\right) H_{SH} \frac{\partial^2 \sigma}{\partial \psi^2}\bigg|_{eq} \right] \end{array} \right\}$$

(7)

Here $f(\psi_{eq}) = \left[ \frac{\partial^2 \sigma}{\partial \psi^2}\bigg|_{eq} + \Gamma \frac{\pi}{2} \frac{\Delta}{\alpha} \left(\frac{2M_S}{\gamma}\right) \gamma H_{SH} sin\psi_{eq} \right]$ and $\psi_{eq}$ is the equilibrium (steady state) magnetization angle of the wall. The domain wall energy density ($\sigma$) is defined as:

$$\sigma = \sigma_0 + M_S H_K \Delta \cos^2 \psi - \pi \Delta M_S H_{DM} \cos \psi \tag{8}$$

where $\sigma_0$ is the domain wall energy density that is just a constant (i.e., not a function of $q$ or $\psi$ ). Note that $\frac{m}{\tau} = \frac{\alpha w t_{FM}}{\Delta} \left(\frac{2M_S}{\gamma}\right)$ gives the friction against the wall motion.

The spin Hall torque tends to rotate the wall magnetization away from the Neel configuration ($\psi_{eq} \sim 0$ or $\pi$) to that of the Bloch configuration ($\psi_{eq} \sim \frac{\pi}{2}$ or $-\frac{\pi}{2}$). When current is applied, one can substitute $\psi_{eq} \sim \frac{\pi}{2}$ or $-\frac{\pi}{2}$ into Supplementary Eq. (6) to obtain the acceleration time ($\tau_A$), as shown in Eq. (2) of the main text. The deceleration time ($\tau_D$) (Eq. (3)) can be evaluated by substituting $\psi_{eq} \sim 0$ or $\pi$ in Supplementary Eq. (6).

One can associate the relaxation time $\tau$ with the effective wall mass $m$ using the relation that derives from Supplementary Eqs. (5-7), i.e. $m_{A(D)} = \frac{2M_S \alpha w t_{FM}}{\gamma \Delta} \tau_{A(D)}$. $m_A$ represents the effective mass when the domain wall is driven by current whereas $m_D$ corresponds to the effective mass when the wall is at rest (i.e. when the current is turned off). Since the proportionality factor that relates $m$ and $\tau$ is a constant, these equations indicate that the effective wall mass is different when the domain wall is driven by current and when it is at rest. Note that $\tau_{A(D)}$ and $m_{A(D)}$ evolve during the transient process and therefore are not constant.

For domain wall tilting, the 1D model predicts that the time it takes to reach the steady state tilting angle ($\tau_\chi$) depends on the wire width and the damping constant. According to Boulle et al.[4], $\tau_\chi$ is expressed as:

$$\tau_\chi = \alpha \frac{M_S w^2}{6\sigma\gamma\Delta} \tag{9}$$

where $\sigma$ is the domain wall energy density at rest (see Ref. 4). The velocity saturates once the tilting is in its equilibrium state: see the black solid lines ($\alpha \sim 0.01$) in Fig. 2(a) and 2(c), which correspond to $\tau_\chi \sim 20$ ns. The 2D pinning and the wall tilting have little effect on the velocity for the short pulses because the time is not enough to develop sizeable amount of tilting.

*Comparison to experimental results*

The current density dependence of $v_{END}$ for the ~5 μm wide wire (Supplementary Fig. 3(b)) is fitted with the 1D model that includes wall tilting and pinning. The model parameters are chosen based on the material parameters of film set B (see Supplementary Table 1). However, $M_S$ used in the calculations is larger than that found in the experiments. *M/V* in Supplementary Table 1 underestimates the saturation magnetization since it includes information of magnetic dead layer. We thus use an intermediate value between *M/V* and the bulk $M_S$ of $Co_{20}Fe_{60}B_{20}$ reported in the literature[16]. To account for the non-zero threshold current density, a one dimensional periodic pinning field[17] $H_{PIN}(q) = \frac{1}{2M_S w t_{FM}} \left(\frac{V_0 \pi}{q_0}\right) sin\left(\frac{\pi}{q_0} q\right)$ has been included in the model to fit the results: here $V_0 = 1.6 \times 10^{-11}$ erg and $q_0 = 7$nm are used. The fitting parameter is the DM exchange constant: the best fit gives $D = 0.24$ erg/cm², which is in agreement with that found from the fitting of $v_{END}$ vs. pulse amplitude (Fig. 1) using Eq. (1). Note that smaller $V_0$ and larger $q_0$ also provide reasonably good agreement for the results shown in Supplementary Fig. 3(b). To reproduce the pulse length dependence of $v_{END}$ shown in Supplementary Fig. 4(a-c), however, small $q_0$ of ~5-10 nm is needed. Such small pinning periodicity is consistent with the amorphous structure of the CoFeB layer. With the $D$ value obtained from the fitting, the equilibrium wall magnetization at rest is pointing close to the $x$ axis, i.e. the wall forms a Neel-like structure[1,18].

The pulse length dependence of $v_{END}$ is calculated using the 1D model with the parameters obtained above. The results are shown by the solid line (pinning is considered) and the dashed line (without pinning) in Supplementary Fig. 4(a-c). In agreement with the experiments, the calculated $v_{END}$ increases with decreasing pulse length for short pulses ($t_P \lesssim 20$ ns). However, the 1D model fails to reproduce the experimental results at longer pulses. Note that $v_{END}$ vs. $t_P$ is nearly identical for the tilted walls and the rigid walls (data not shown). In the 1D model, the wall tilting angle is underestimated due to the 1D nature of the pinning and consequently, the tilting has little effect on $v_{END}$. We find that the velocity reduction for longer pulses is only well reproduced when full micromagnetic simulations with realistic 2D disorder are considered.

*Validity of the linearized equation of motion*

The wall angle $\psi$ changes from 0 (or $\pi$) to $\pm\pi/2$ and vice versa when the current is tuned on and off. Thus the linearization process used to obtain the relaxation times (Eqs. (2) and (3)) needs justification. To study this, we have numerically calculated the domain wall velocity and extracted the relaxation times ($\tau_A$ and $\tau_D$) using the solutions of the linearized model. Supplementary Fig. 6(a,b) show the instantaneous velocity as a function of time when a 100 ns long current pulse is applied. The parameters used are similar to those described in Fig. 2 (with Gilbert damping α=0.05). For simplicity, we assume the tilt angle $\chi$ to be zero here. The acceleration and deceleration times are obtained by fitting the velocity vs. time using the following exponential function, the solution of the linearized equation (Supplementary Eq. (4)). With $v(t) = \frac{\partial q(t)}{\partial t}$, the solution takes the form:

$$v(t) = \begin{cases} v_1 \left[1 - \exp\left(-\frac{t}{\tau_A}\right)\right] & \text{for } 0 \leq t < t_P \\ v_2 \exp\left(-\frac{t-t_P}{\tau_D}\right) & \text{for } t \geq t_P \end{cases} \quad (10)$$

where $v_{1(2)}$ are the fitting parameters and $t_P$ is the current pulse length. The boundary condition at $t = t_P$ suggest that $v_1 = v_2$ (in the text, we use $v_1 = v_2 = v_D$). Here, for the purpose of fitting, we use two different parameters $v_{1(2)}$ for the reason described below.

As the time variation of the wall angle is the main source of the relaxation effects, we have calculated and fitted $\psi$ vs. time using similar exponential functions (i.e. replace $v_{1(2)}$ with $\psi_{1(2)}$): the calculated and fitted curves are shown in Supplementary Fig. 6(c,d). The left and right panels show calculation results using different current densities. Note that the equation of motion (Supplementary Eq. (4)) and its solution (Supplementary Eq. (10)), i.e. the exponential function, are valid only when the wall angle $\psi$ is close to $\pm\pi/2$ when the current is on and 0 (or $\pi$) when it is off. We therefore limit the fitting range to which the exponential function can be applied: to a time range in which deviation of $\psi$ from its equilibrium value is less than ~20 deg. This is why we have to define the amplitudes of the exponential function ($v_{1(2)}$ and $\psi_{1(2)}$) separately when the current is on and off.

First, from the fitting, we obtain the saturation velocity ($v_D$) and the corresponding equilibrium wall angle (i.e. $\psi_{eq}$) when current is applied. $v_D$ and $\psi_{eq}$ are equivalent to, respectively, $v_1$ and $\psi_1$ in Supplementary Eq. (10). These quantities are plotted in Supplementary Figs. 6(e) and 6(f) using the solid symbols. (As a guide to the eye, the solid line in Supplementary Fig. 6(e) shows numerically calculated saturation velocity at the end of the current pulse ($t=100$ ns).) Supplementary Figs. 6(e) and 6(f) show that $\psi_{eq}$ decreases with decreasing current density, resulting in a smaller velocity at lower current. In the parameter set used here, a considerably decrease in $\psi_{eq}$ and $v_D$ occurs when the current density is smaller than $\sim 0.2 \times 10^8$ A/cm$^2$.

The corresponding acceleration ($\tau_A$) and deceleration ($\tau_D$) times are shown in Supplementary Figs. 6(g) and 6(h). We show $\tau_A$ and $\tau_D$ obtained by fitting the velocity vs. time (black squares) and $\psi$ vs. time (red circles) and compare those to the values calculated using Eqs. (2) and (3) (blue solid line). We find that the numerical calculations and the analytical solutions of the acceleration time $\tau_A$ are in good agreement even for small current densities at which $\psi_{eq}$ is much smaller than $\pi/2$. These calculations show that the estimation of $\tau_A$ using Eq. (2) is valid at smaller current although its derivation assumes $\psi_{eq} \approx \frac{\pi}{2}$.

The numerical calculations of the deceleration time (Supplementary Fig. 6(h), solid symbols) show that $\tau_D$ varies little with the current density. This is in good agreement with Eq. (3), which dictates that $\tau_D$ is constant against the current density. The numerical calculations of the deceleration time are ~10-20% smaller than the analytical estimate. From these results, we consider the expressions given in Eqs. (2) and (3) provide good estimates of the relaxation times.

The difficulty in fitting the relaxation process arises since the wall mass, or the relaxation time, continues to evolve during the transient processes. Under such condition, it is not appropriate to use a single relaxation time to describe the process. We have therefore limited the fitting range to estimate a relaxation time that more or less describes the equilibrium state. Ideally, to describe the relaxation process, one would need to use a relaxation time that is a weighted average of the processes involved.

*Numerical evaluation of the relaxation times*

As the velocity or the wall angle cannot be fitted well with an exponential function with a constant relaxation time, we have computed the relaxation times numerically using the calculation results of the 1D model. To illustrate how the relaxation times are obtained numerically, we show in Supplementary Fig. 7(a) the temporal evolution of the wall velocity when a current pulse is applied. The results are similar to that presented in Fig. 2(a), solid line. The maximum velocity ($v_{\max}$) and the velocity at the end of the pulse, i.e. the terminal velocity ($v(t_P)$) are defined schematically in Supplementary Fig. 7(a). The acceleration time ($\tau_A$) is obtained by calculating the time needed to reach $\frac{2}{3}v_{\max}$ after the pulse is turned on. The deceleration time ($\tau_D$) is estimated by the time it takes to reach $\frac{1}{3}v(t_P)$ after the pulse is turned off. Although the relaxation times obtained in such a way quantitatively differ from those calculated using the linearized solutions (Eqs. (2) and (3)), the former provides a qualitative view of how the relaxation times depend on key material parameters.

The numerically calculated relaxation times are displayed in Supplementary Figs. 7(b-d) as a function of current density and in Supplementary Figs. 7(e-g) as a function of Gilbert damping, spin Hall angle and the DM exchange constant. The difference in $\tau_A$ and $\tau_D$ is apparent when the damping is small and when the DM exchange constant is small such that $\frac{\pi}{2}H_{\mathrm{DM}}$ approaches the domain wall anisotropy field ($H_K$). These results qualitatively support the relaxation times (Eqs .(2) and (3)) obtained using the linear approximation of the 1D model.

## Supplementary Note 3

**Full micromagnetic simulations**

In order to further support the experimental observations and the 1D model calculations, full micromagnetic (μM) simulations have been performed by solving the Landau Lifshitz Gilbert equation augmented with the damping-like component of the spin Hall torque:

$$\frac{\partial \vec{m}}{\partial t} = -\gamma \vec{m} \times (\vec{H}_{\text{eff}} + \vec{H}_{\text{th}}) + \alpha \vec{m} \times \frac{\partial \vec{m}}{\partial t} + \gamma \frac{\hbar \theta_{\text{SH}} J(t)}{2 e M_S L_z} \vec{m} \times (\vec{\sigma} \times \vec{m}) \qquad (11)$$

where the effective field $\vec{H}_{\text{eff}}$ includes exchange, magnetostatic, magnetocristalline anisotropy (i.e. uniaxial perpendicular magnetic anisotropy) and Dzyaloshinskii-Moriya interactions. $\vec{H}_{\text{th}}$ is the thermal field and $\vec{\sigma} = \vec{u}_y$ is the polarization of the spin current (see Ref. [19] for numerical details) entering the magnetic layer. The material parameters are the same with those used for the 1D model: $M_S = 1100$ emu/cm$^3$, $A = 1.5 \times 10^{-6}$ erg/cm, $K_{\text{EFF}} \sim 3.2 \times 10^6$ erg/cm$^3$, $\theta_{\text{SH}} = -0.21$, $\alpha = 0.05$ and $D = 0.24$ erg/cm$^2$.

In order to take into account the effects of disorder due to imperfections and defects in a more realistic way than that of the 1D model, we assume the easy axis anisotropy direction is distributed among a length scale defined by a "grain" size. The grains vary in size taking an average diameter of $D_G = 30$ nm. The direction of the uniaxial anisotropy of each grain is mainly directed along the perpendicular direction (z-axis) but with a small in-plane component which is randomly generated over the grains. The maximum percentage of the in-plane component of the uniaxial anisotropy unit vector is varied from 10% to 15% ($0.10 \leq \varepsilon \leq 0.15$). In this work, we have computed the domain wall velocity as a function of pulse length for five different grain patterns (A-E) generated randomly and the average velocity are compared to the experimental results shown in Fig. 5(a-c).

In order to evaluate the influence of the wall tilting, two strips with two different widths were studied numerically using 2D micromagnetic simulations: $w = 1536\ nm$ and $w = 4997\ nm$. Note that the latter is the same as that of the experimental wire studied in the main text. The strips are discretized using a finite difference scheme with cells composed of $3nm \times 3nm \times 1nm$: the thickness of the cell is the same with that of the CoFeB strip ($t_{FM} = 1nm$). A micromagnetic study using the real dimensions of the experimental samples (~30-40 μm long wires) is not possible due to computer memory limitations. Therefore, the length of the strips considered in the modeling is $l = 12.3\ \mu m$. Similar to the experiments, two domain walls are placed in the strips and the current-driven motion of domain walls is evaluated. The quasi-static velocity $v_{END}$ is estimated by dividing the total distance the domain wall traveled both during and after the current pulse application with the pulse length.

We first focus on the strip with a width of $w = 1536\ nm$ and study the effect of the degree of disorder ($\varepsilon$) on $v_{END}$. Supplementary Fig. 8 shows the simulated $v_{END}$ as a function of the pulse length $t_P$ for grain pattern A with three degrees of disorder: $\varepsilon = 0.10, 0.12, 0.15$. The current density is fixed to $J = 0.8 \times 10^8$ A/cm$^2$, a condition that corresponds to that of Supplementary Fig. 4(c). The $t_P$ dependence of $v_{END}$ is similar for the three degrees of disorder evaluated. In terms of quantitative agreement with the experiments (black solid circles in Supplementary Fig. 8) the best fit is found for the case with $\varepsilon = 0.12$. Based on this agreement, the degree of disorder is fixed to $\varepsilon = 0.12$ from hereafter.

The effects of grain pattern and temperature on $v_{END}$ are presented in Supplementary Fig. 9. Similar results are obtained for different grain patterns and with different temperatures, i.e. zero and room temperature. The simulations are in good agreement with the experimental results. Snapshots of the magnetic contrast, before and after the current pulse application, are shown in Supplementary Fig. 9(d-f). For larger $t_P$, in contrast to the 1D model calculations, the tilting angle is non-zero even after the current pulse is turned off. In addition, we find that the tilting angle during the current pulse is larger when the 2D pinning is introduced compared to that estimated using the 1D model. Thus for longer pulses, where the tilting becomes more significant, the velocity reduction is larger in the simulations and thus accounts for the gradual reduction of $v_{END}$ with increasing $t_P$. For short pulses ($t_P \lesssim 20$ ns), the inertia effect determines the enhancement in $v_{END}$, as predicted by the 1D model.

As the domain wall tilting scales with the strip width $w$, the strip width influences the time scale of domain wall tilting (see Supplementary Eq. 10) and therefore it can modify the domain wall velocity. In order to evaluate this effect a second micromagnetic study was carried out using the same strip width as in the experimental measurements ($w \sim 5\ \mu m$). The results of these simulations are shown in Supplementary Fig. 10 in comparison to the experimental results. As it is clearly shown, the results for the wider strip are in very good agreement with the experimental data and they exhibit similar trend with the previous simulations for narrow wires in Supplementary Fig. 9. Agreement with experimental results is slightly better for simulations with the wider wires. Note that the increase in the velocity at shorter pulses is a little more abrupt for the narrower wire (Supplementary Fig. 9) compared to that of the wider wire (Supplementary Fig. 10), which is due to the wire width dependent time scale of domain wall tilting.

These micromagnetic simulations corroborate the experimental results and the interpretation based on the 1D model discussed in the main text. The 1D model description is valid for short current pulses with $t_P \lesssim 10 - 20$ ns. However, as the pulse length increases, the model fails to provide a quantitative agreement with the experimental results. The reasons behind this are described as follows. In the framework of the 1D model, the pinning is introduced as a one dimensional space-dependent effective field defined by a given energy barrier and a period. This pinning field is purely 1D (only depends on the *x* coordinate), and therefore it cannot capture the 2D pinning present in real samples. As $t_P$ increases, the domain wall tilting increases, however its degree is larger when a 2D pinning is assumed. Since the velocity becomes smaller as the tilting increases, the velocity reduction is larger in the simulations (compared to the 1D model calculations) due to the 2D pinning that gives rise to larger tilting.

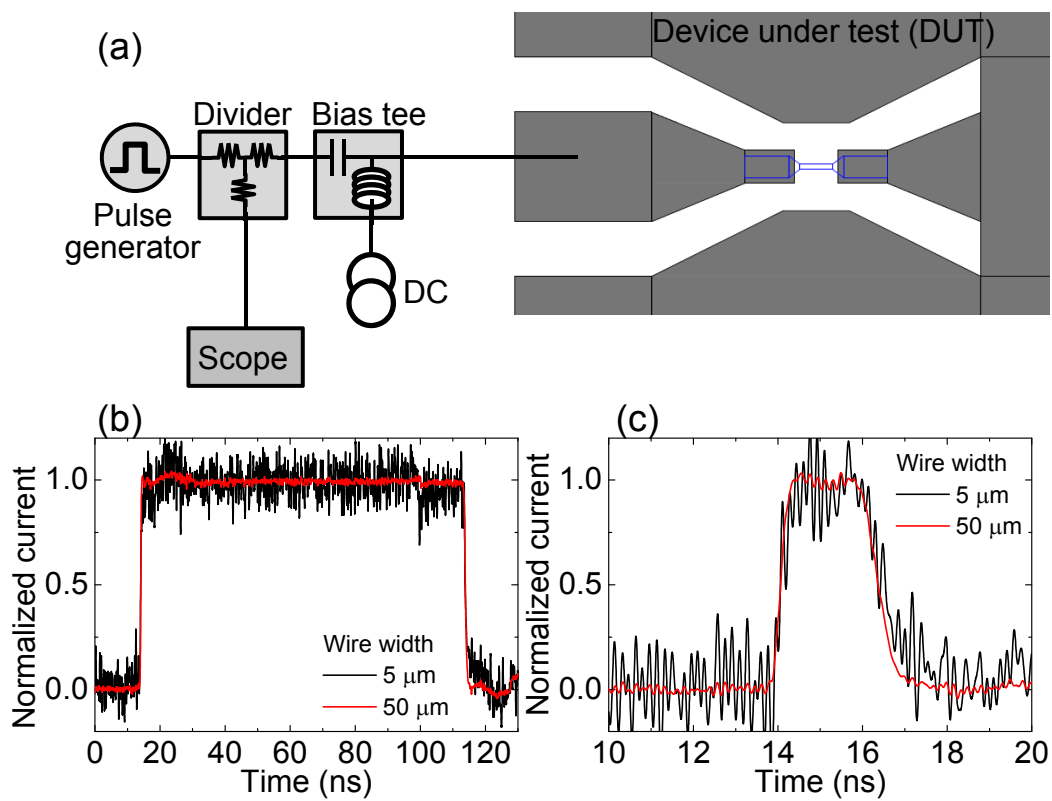

**Supplementary Figure 1. Pulse transmission characteristics of a typical device under investigation.** (a) Schematic illustration of the measurement setup. (b,c) Measured current that flows into the device when a pulse is applied from the pulse generator for two different devices, $w$~5 μm and ~50 μm. The current is estimated using the time domain reflection measurements. The pulse length is ~100 ns (b) and ~2.1 ns (c). Results are from film set A.

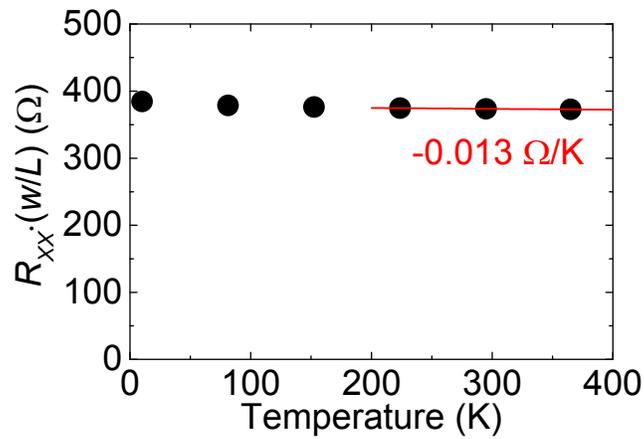

**Supplementary Figure 2. Temperature dependence of the device resistance.** (a) The resistance ($R_{XX}$) of a wire made from Sub.|W(~3.1)/Co$_{20}$Fe$_{60}$B$_{20}$(1)/MgO(2)/Ta(1) is measured as a function of measurement temperature ($T$). The resistance is normalized by the length ($L$) and width ($w$) of the wire. The resistance hardly changes with temperature: the slope of $R_{XX} \cdot (w/L)$ vs. $T$ is ~−0.013 W/K. Results are from film set B.

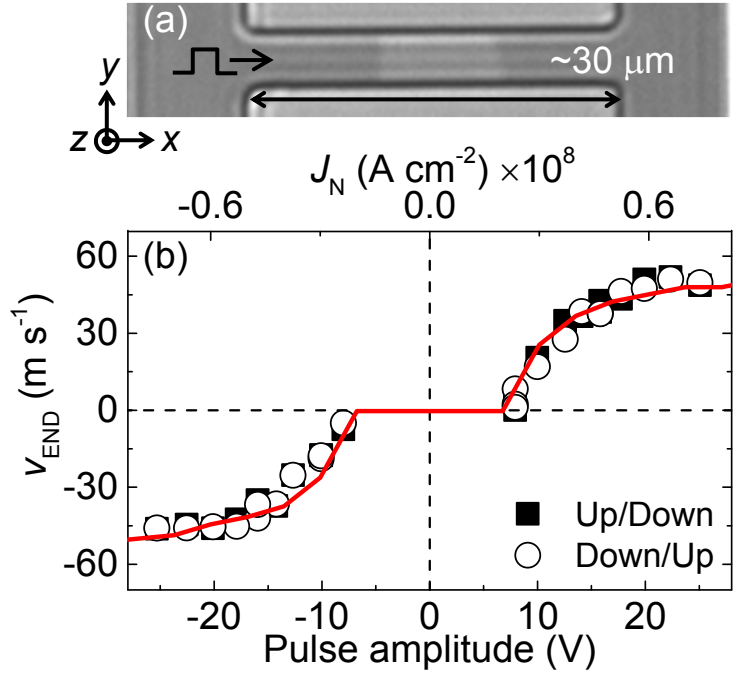

**Supplementary Figure 3. Fitting of velocity vs. current density using the 1D model.** (a) Optical (Kerr) microscopy image of the device used to study domain wall motion. (b) Pulse amplitude dependence of the quasi-static velocity ($v_{END}$) for a fixed pulse length ($t_P = 10$ ns). The corresponding current density that flows through the W underlayer is shown in the top axis. The open and solid symbols show the velocity for ↓↑ walls and ↑↓ walls, respectively. Results are from film set B, wire width is ~5 µm. The red solid line represents fitting with the 1D model that takes into account domain wall tilting and pinning. Parameters used: $M_S = 1100$ emu cm$^{-3}$, $K_{EFF} = 3.2 \times 10^6$ erg cm$^{-3}$, $\Delta = \sqrt{A/K_{EFF}} \sim 6.8$ nm ($A = 1.5 \times 10^{-6}$ erg cm$^{-1}$), $\theta_{SH} = -0.21$, $D = 0.24$ erg cm$^{-2}$, $\alpha = 0.05$ and the wire width $w=5$ µm. A 1D pinning $H_{PIN}(q) = \frac{1}{2M_S w t_{FM}} \left(\frac{V_0 \pi}{q_0}\right) \sin\left(\frac{\pi}{q_0} q\right)$ that accounts for local imperfections is introduced with $V_0=1.6 \times 10^{-11}$ erg and $q_0=7$ nm.

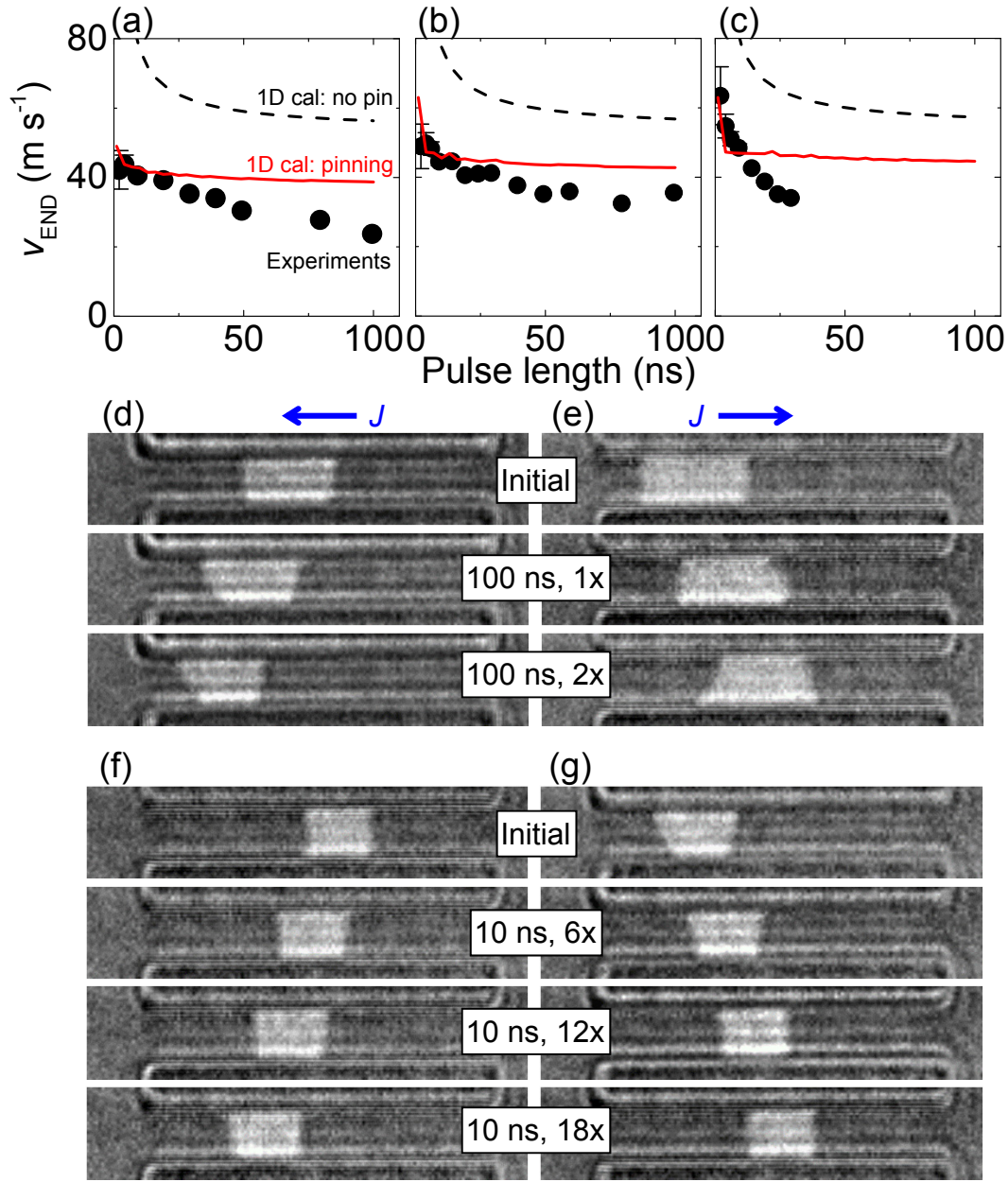

**Supplementary Figure 4. Current driven domain wall motion in ~5 μm wide wires.** (a-c) Experimentally measured quasi-static velocity ($v_{END}$, black circles) as a function of pulse length for three different pulse amplitudes: (a) 16 V ($J \sim 0.5 \times 10^8$ A cm$^{-2}$) (b) 20 V ($J \sim 0.6 \times 10^8$ A cm$^{-2}$) and (c) 25 V ($J \sim 0.8 \times 10^8$ A cm$^{-2}$). The errors bars represent variation in $v_{END}$ due to the uncertainty in the pulse length. Experimental results are the same with those shown in Fig. 5(a-c). The red solid and black dashed lines show $v_{END}$ calculated using the 1D model with and without pinning, respectively. Parameters used are the same with those shown in Supplementary Fig. 3 caption. (d-g) Sequences of Kerr images showing the successive wall motion after application of current pulses described in the legend: (d) $t_P = 100$ ns, $J<0$, (e) $t_P = 100$ ns, $J>0$, (f) $t_P = 10$ ns, $J<0$ and (g) $t_P = 10$ ns, $J>0$. All results are from film set B, wire width is ~5 μm.

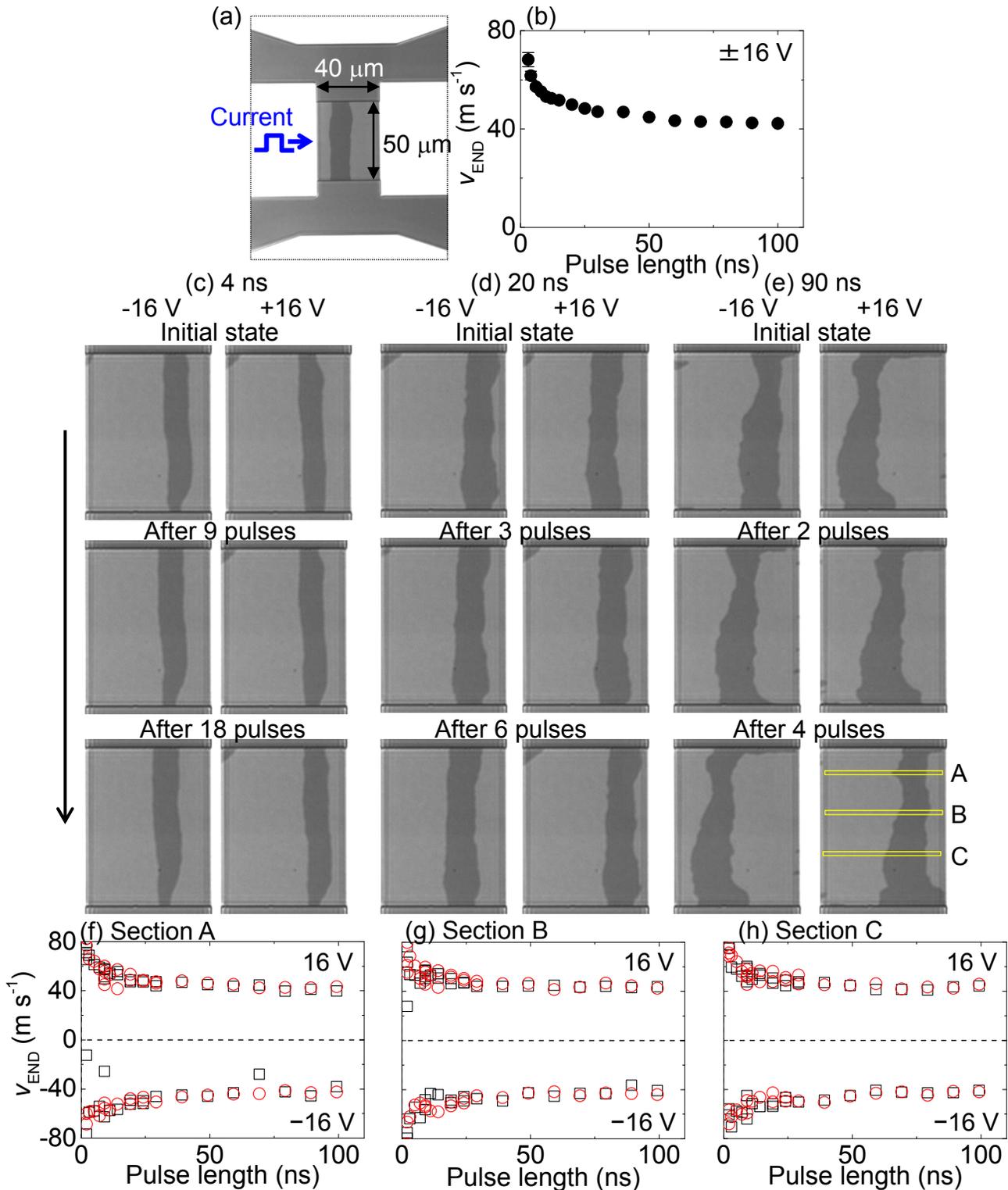

**Supplementary Figure 5. Current driven domain wall motion in ~50 μm wide wires.** (a) Representative Kerr microscopy image of the ~50 μm wide wire used to study domain wall motion. (b) Average of the quasi-static velocity ($v_{END}$) obtained from the three sections A, B, C denoted by the yellow rectangles in (e). The error bars represent standard deviation of the velocity estimated in the three sections. (c-e) Sequences of Kerr images showing the successive wall motion after application of current pulses described in the legend. The left and right panels show images when -16 V and +16 V pulses are applied, respectively. The pulse length is (c) 4 ns, (d) 20 ns and (e) 90 ns. (f-h) $v_{END}$ as a function of pulse length obtained from the three sections A (f), B (g) and C (h) denoted by the yellow rectangles in (e). The pulse amplitude is fixed to ±16 V. Black squares and red circles represent $v_{END}$ for ↑↓ and ↓↑ walls. The film structure is Sub.|W(~3.6)/Co$_{20}$Fe$_{60}$B$_{20}$(1)/MgO(2)/Ta(1). Results are from film set A, wire width is ~50 μm.

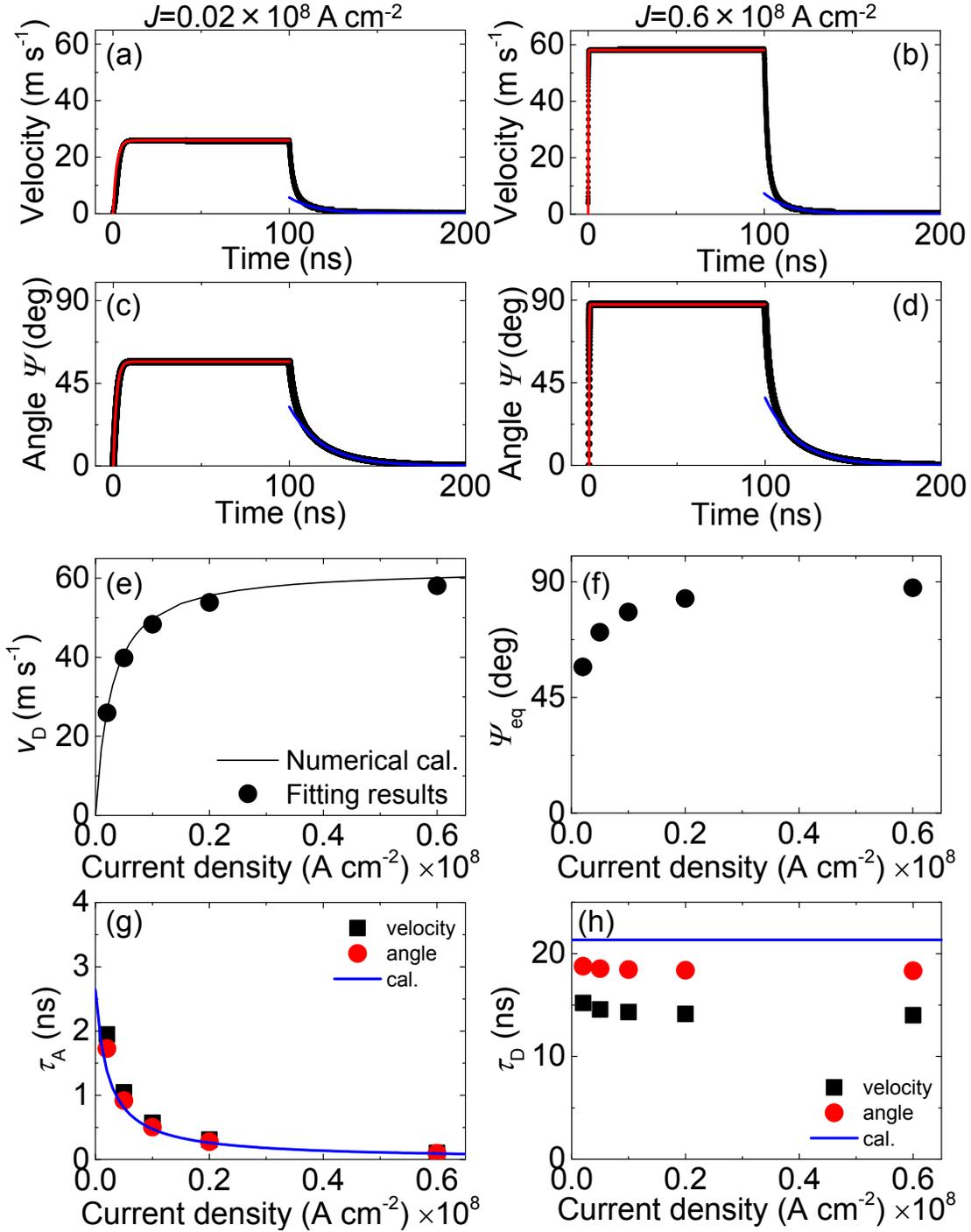

**Supplementary Figure 6. Analyses of the linearized 1D model.** (a-d) Instantaneous DW velocity $v(t)$ (a,b) and the wall magnetization angle $\psi(t)$ (c,d) for a fixed current density of $J = 0.02 \times 10^8$ A cm$^{-2}$ (a,c) and $J = 0.6 \times 10^8$ A cm$^{-2}$ (b,d) flowing through the heavy metal layer. The current pulse length is ($t_P$) is 100 ns. Fit to data in appropriate ranges using Supplementary Eq. (10) are shown by the red and blue solid lines. (e-h) Current density $J$ dependence of saturation velocity (e), the equilibrium wall angle (f), the acceleration time (g) and the deceleration time (h). Results are obtained by the fitting process described in (a-d). The solid line in (e) shows numerically calculated saturation velocity at the end of the current pulse ($t$=100 ns). The blue solid line in (g) and (h) are the analytical solutions provided in Eqs. (2) and (3), respectively. Parameters used: $M_S = 1100$ emu cm$^{-3}$, $K_{\text{EFF}} = 3.0 \times 10^6$ erg cm$^{-3}$, $\Delta = \sqrt{A/K_{\text{EFF}}} \sim 7.0$ nm ($A = 1.5 \times 10^{-6}$ erg cm$^{-1}$), $\theta_{\text{SH}} = -0.21$, $D = 0.24$ erg cm$^{-2}$, $\alpha = 0.05$ and $w$=5 μm.

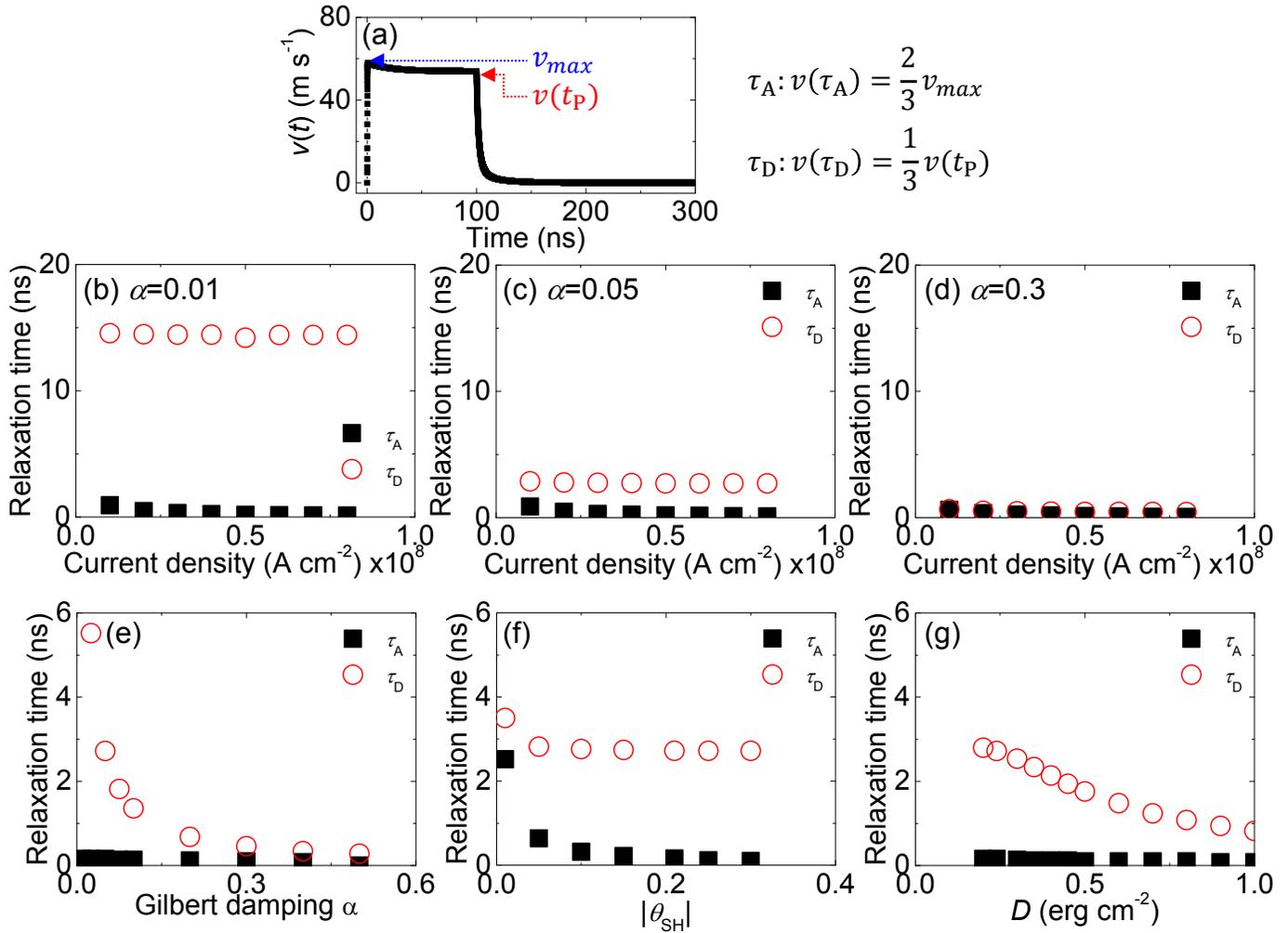

**Supplementary Figure 7. Numerical calculations of the relaxation times using the 1D model.** (a) Instantaneous DW velocity $v(t)$ when a pulse with current density of $J = 0.6 \times 10^8$ A cm$^{-2}$ is applied to the heavy metal layer. The current pulse length is ($t_P$) is 100 ns. The maximum velocity ($v_{max}$) and the velocity at the end of the pulse, i.e. the terminal velocity ($v(t_P)$) are illustrated schematically. (b-g) Acceleration ($\tau_A$) and deceleration ($\tau_D$) times numerically computed using the definition described in (a). $\tau_A$ and $\tau_D$ are plotted as a function of current density (b-d), Gilbert damping constant $\alpha$ (e), spin Hall angle $\theta_{SH}$ (f) and the DM exchange constant $D$ (g). The Gilbert damping constant $\alpha$ is varied for the plots shown in (b-d): $\alpha$ =0.01 (b), 0.05 (c) and 0.3 (d). The parameters used unless specified are: $M_S = 1100$ emu cm$^{-3}$, $K_{EFF} = 3.2 \times 10^6$ erg cm$^{-3}$, $\Delta = \sqrt{A/K_{EFF}} \sim 6.8$ nm ($A = 1.5 \times 10^{-6}$ erg cm$^{-1}$), $\theta_{SH} = -0.21$, $D = 0.24$ erg cm$^{-2}$, $\alpha = 0.05$ and $w=5$ μm.

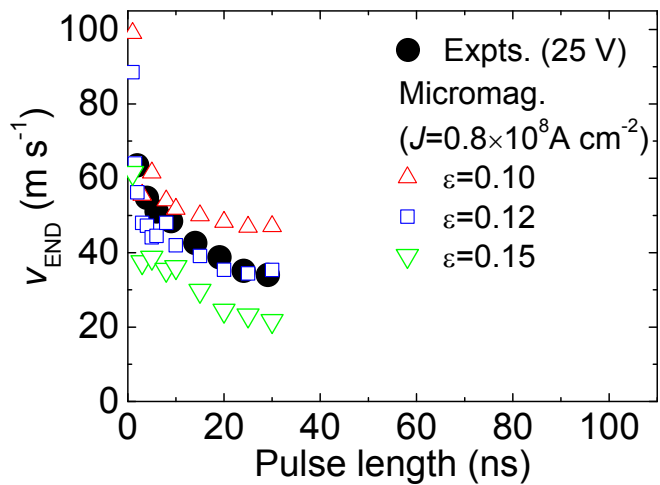

**Supplementary Figure 8. Micromagnetic simulations of the quasi-static velocity: degree of disorder dependence.** Quasi-static velocity ($v_{END}$) as function of the pulse length under current pulses of fixed amplitude $J = 0.8 \times 10^8 \text{A cm}^{-2}$. Experimental results (solid circles, same with those shown in Fig. 5(c)) are compared to the full micromagnetic predictions considering different degrees of disorder and grain pattern A. The width of the wire ($w$) is ~5 μm for the experiments (film set B) and $w$ is set to 1.5 μm for the simulations.

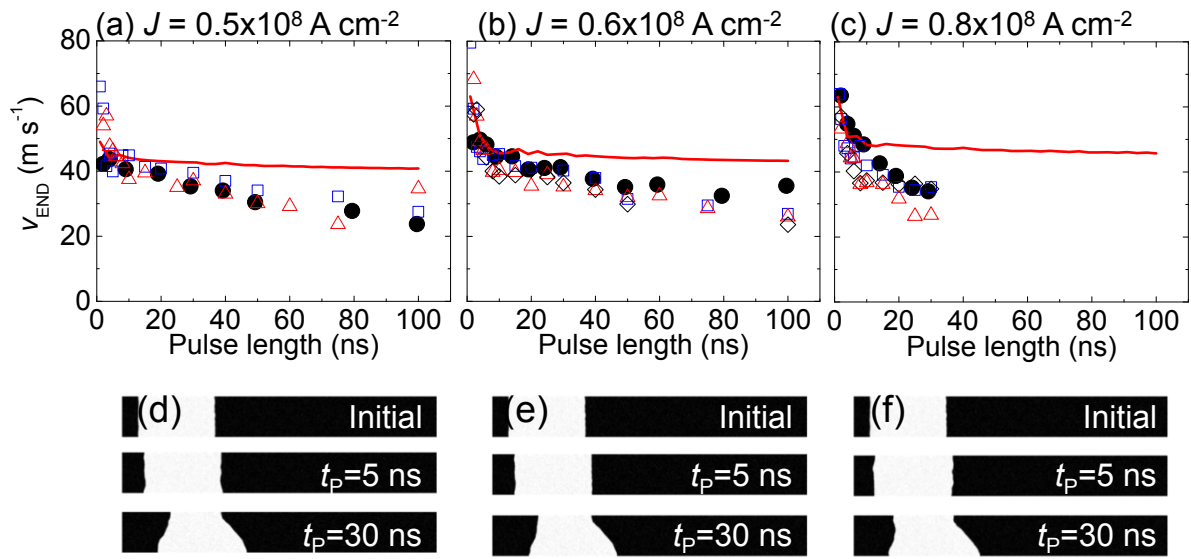

**Supplementary Figure 9. Comparison of calculated quasi-static velocity using 1D model and micromagnetic simulations.** (a-c) Quasi-static velocity ($v_{END}$) as function of the pulse length under current pulses of fixed amplitude: (a) $J = 0.5 \times 10^8$ A cm$^{-2}$, (b) $J = 0.6 \times 10^8$ A cm$^{-2}$ and (c) $J = 0.8 \times 10^8$ A cm$^{-2}$. Experimental results (black circles, same with those shown in Fig. 5(a-c)) are compared to full micromagnetic simulations (open symbols). The degree of disorder in the micromagnetic simulations is fixed to $\varepsilon = 0.12$. Blue, red and black open symbols correspond to micromagnetic results using the following conditions: blue squares are for grain pattern A with $T$=0 K, red triangles are for grain pattern A with $T$=300 K and black diamonds are for grain pattern B with $T$=0 K. Grain patterns A and B are randomly generated. (d-f) Micromagnetic snapshots of the initial state and states long after application of current pulses ($t_P = 5$ and 30 ns). The width of the wire ($w$) is ~5 μm for the experiments (film set B) and $w$ is set to 1.5 μm for the simulations.

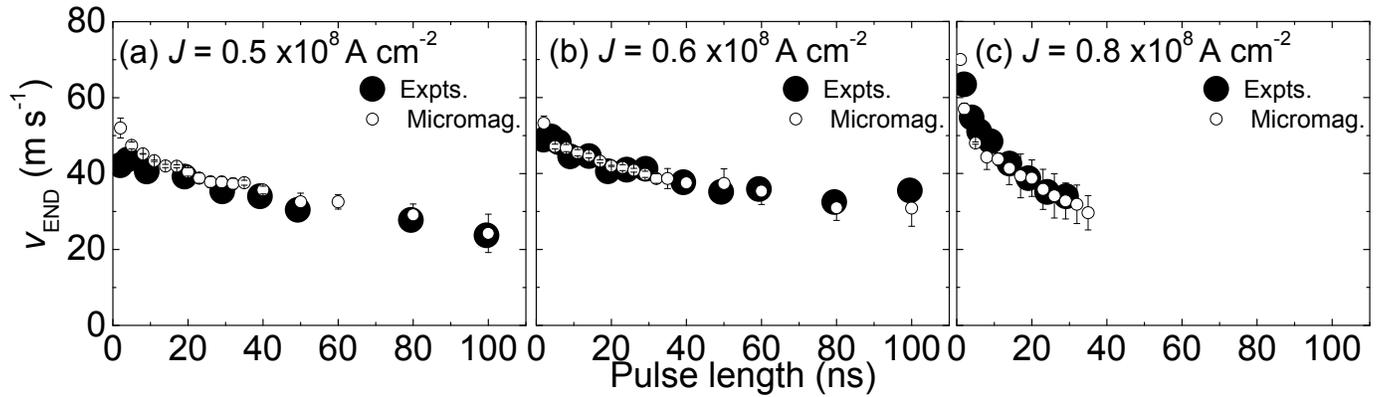

**Supplementary Figure 10**. **Micromagnetic simulations of the quasi-static velocity: current density dependence** Quasi-static velocity ($v_{END}$) as function of the pulse length under current pulses of fixed amplitude. Experimental results (big solid circles, same with those shown in Fig. 5(a-c)) are compared to full micromagnetic simulations (small open circles) with wire width set close to the experiments. (a) $J = 0.5 \times 10^8$ A cm$^{-2}$, (b) $J = 0.6 \times 10^8$ A cm$^{-2}$ and (c) $J = 0.8 \times 10^8$ A cm$^{-2}$. The degree of disorder in the micromagnetic simulations is fixed to $\varepsilon = 0.12$ for grain pattern A with $T$=300 K. A grain size of 20 nm was considered for these micromagnetic results. The width of the wire ($w$) is 5 μm for both the experiments (film set B) and the simulations. The error bars of the simulations represent five independent calculations using different thermal noise patterns.

**Supplementary Table 1. Magnetic and transport properties of film sets A and B.** The film structure is Sub./W($d$)/Co$_{20}$Fe$_{60}$B$_{20}$(1)/MgO(2)/Ta(1) (units in nanometers). Resistivity ($\rho_N$), absolute value of the spin Hall angle ($|\theta_{SH}|$), spin diffusion length ($\lambda_N$), saturation magnetization ($M_S$) and the effective magnetic anisotropy energy ($K_{EFF}$) are listed. For $M_S$ and $K_{EFF}$ values are taken from films with the W layer thickness $d\sim3$ nm.

| Films | $\rho_N$ μΩcm | $|\theta_{SH}|$ | $\lambda_N$ nm | $M_S$ [a] emu cm$^{-3}$ | $K_{EFF}$ [a] erg cm$^{-3}$ |
|---|---|---|---|---|---|
| Set A | 150 | 0.24 | 1.1 | 790 | $2.9 \times 10^6$ |
| Set B | 125 | 0.23 | 1.3 | 780 | $3.2 \times 10^6$ |

[a] Values when $d\sim3$ nm